\documentclass[aps,prx,amsmath,reprint,floatfix]{revtex4-2}
\usepackage{graphicx}
\usepackage{bm,bbm}
\usepackage[urlcolor=blue]{hyperref}
\hypersetup{colorlinks=true,allcolors=blue}
\usepackage{amssymb} 

\newcommand{\E}[1]{\left\langle{ #1}\right\rangle}
\newcommand{\Es}[1]{\E{ #1}_{\rm s}}
\newcommand{\rmd}{d}
\newcommand{\rme}{\mathrm{e}}

\newcommand{\f}[1]{\mathbf{#1}}
\newcommand{\x}{\f x}
\newcommand{\y}{\f y}
\newcommand{\z}{\f z}

\newcommand{\abs}[1]{\left\lvert #1 \right\rvert}
\newcommand{\norm}[1]{\left\lvert\left\lvert #1 \right\rvert\right\rvert}

\newcommand{\bsig}{\boldsymbol{\sigma}}

\newcommand{\beps}{{\boldsymbol{\varepsilon}}}

\newcommand{\bj}{\hat{\f j}}
\newcommand{\ps}{p_{\rm s}}
\newcommand{\js}{{\f j}_{\rm s}}

\newcommand{\revj}{{\hat{\f j}}^\ddag}
\newcommand{\integrals}{\hat{\mathcal{I}}}

\usepackage{color}\usepackage{xcolor}
\colorlet{mylinkcolor}{blue!66!black!80}
\newcommand{\blue}[1]{{\color{mylinkcolor}#1}}
\renewcommand{\blue}[1]{{#1}}

\newcommand{\finalchange}[1]{{\color{black}#1}}

\begin{document}
\title{Coarse Graining Empirical Densities and Currents in Continuous-Space Steady States}
\author{Cai Dieball}
\author{Alja\v{z} Godec}
\email{agodec@mpinat.mpg.de}
\affiliation{Mathematical bioPhysics Group, Max Planck Institute for Multidisciplinary Sciences, Am Fa\ss berg 11, 37077 G\"ottingen}
\begin{abstract}
We present the conceptual and technical
background required to describe and understand the correlations and
fluctuations 
of the empirical density and current of steady-state
diffusion processes on all time scales --- observables central to 
statistical mechanics and thermodynamics on the level of individual
trajectories. We focus on the important and non-trivial effect of a spatial coarse graining. Making use of a generalized
time-reversal symmetry we provide deeper insight about the physical
meaning of fluctuations of the coarse-grained empirical density and
current, and explain why a systematic variation of the
coarse-graining scale offers an efficient method to infer bounds on
a system's dissipation. Moreover, we discuss emerging symmetries in the
statistics of the empirical density and current, and the statistics in the \finalchange{central-limit} regime. 
More broadly our work promotes the application of stochastic
calculus as a powerful direct alternative to Feynman-Kac theory and path-integral
methods.
\end{abstract}
\maketitle

\section{Introduction}
A non-vanishing probability current \cite{Bodineau2004PRL,Bertini2005PRL,Zia2007JSMTE,Maes2008PA,Gingrich2016PRL,Maes2008EEL,Barato2015JSP,Baiesi2009JSP,Chernyak2009JSP,Bertini2015RMP,Pietzonka2016PRE,Gingrich2017PRL,Barato2018NJP,Kaiser2018JSP,Dechant2018JSMTE,Battle2016S,Li2019NC} and
entropy production
\cite{Roldan2010PRL,Dabelow2019PRX,Pigolotti2017PRL,Seifert2012RPP,Seifert2005PRL,Esposito2010PRE,VandenBroeck2010PRE,Vaikuntanathan2009EEL,Qian2013JMP,Lapolla2020PRL}
are the hallmarks of non-equilibrium, manifested as transients
during relaxation
\cite{Vaikuntanathan2009EEL,Maes2011PRL,Qian2013JMP,Maes2017PRL,Shiraishi2019PRL,Lapolla2020PRL,Koyuk2020PRL}
or in
non-equilibrium, current-carrying
steady states
\cite{Jiang2004,Maes2008PA,Maes2008EEL,Gingrich2016PRL,Seifert2010EEL,Barato2015PRL}. Genuinely
irreversible, detailed balance violating
dynamics
emerge in the presence of non-conservative
forces (e.g.\ shear or rotational flow)
\cite{Schroeder2005PRL,Harasim2013PRL,Gerashchenko2006PRL,AlexanderKatz2006PRL}
or active driving in living matter fueled by ATP-hydrolysis
\cite{Qian2000PRL,Qian2007ARPC,Toyabe2010PRL,Marchetti2013RMP,Fakhri2014S,Fodor2016PRL,Gladrow2016PRL,Battle2016S,Gnesotto2018RPP}. Such
systems are typically small and ``soft'', and thus subject to large
thermal fluctuations.
Single-molecule 
\cite{Gladrow2016PRL,Gnesotto2018RPP,Ritort2006JPCM,Greenleaf2007ARBBS,Moffitt2008ARB}
and particle-tracking \cite{Burov2011PCCP} experiments probe dynamical
processes on the level of individual, stochastic trajectories. These
are typically analyzed within the framework
of ``time-average statistical mechanics'' \cite{Burov2011PCCP,Lapolla2020PRR,Qian_PRE,Horowitz2019NP,Gingrich2016PRL,Gingrich2017JPAMT,Dechant2021PRR,Dechant2021PRX}, i.e.\ 
by averaging along
individual finite realizations
yielding random quantities with nontrivial statistics. 

Ergodic steady states are characterized by the (invariant) steady-state
density $\ps(\x)$ and a steady-state probability current $\js(\x)$ in systems
with a broken detailed balance. One can equivalently infer $\ps(\x)$ and
$\js(\x)$  from an
ensemble of statistically independent trajectories of an ergodic process, or from an
individual but very long (i.e.\ ergodically long \footnote{An ergodic
time scale is longer than any
correlation time in the system.}) trajectory. 
To infer $\ps(\x)$ and $\js(\x)$ from individual sample paths one uses estimators
that are called the \emph{empirical density} and \emph{empirical
current}, respectively, defined as
\begin{align} 
\overline{\rho^U_\x}(t)&\equiv \frac{1}{t}\int_0^t
U^h_\x(\x_\tau)\rmd\tau\nonumber\\
\overline{\f J^U_\x}(t)&\equiv \frac{1}{t}\int_{\tau=0}^{\tau=t}U^h_\x(\x_\tau)\circ\rmd\x_\tau,
\label{def_current}
\end{align}
where $U^h_\x(\z)$ is a ``window function'' around a point $\x$ with a characteristic
scale $h$ \cite{AccompanyingLetter} and $\circ\,\rmd\x_\tau$ denotes the Stratonovich integral, which both will
be specified more precisely below. Notably, the Stratonovich integration $\circ\rmd\x_\tau$
in Eq.~\eqref{def_current} is the correct way to make
sense of the expression $"\dot\x_\tau\rmd\tau"$, which is ill-defined since for any $\tau$ with probability one
$\abs{\dot\x_\tau}=\infty$ for overdamped Langevin dynamics
\cite{Durrett_Stoch}. Because
$(\x_\tau)_{0\le \tau\le t}$ is random, $\overline{\rho^U_\x}(t)$ and $\overline{\f
  J^U_\x}(t)$ are fluctuating quantities. Notably, the empirical
density and current are typically defined with a delta function,
i.e.\ with $U^{h\to 0}_\x(\z)=\delta(\x-\z)$ \cite{Maes2008PA,Touchette2009PR,Kusuoka2009PTRF,Chetrite2013PRL,Chetrite2014AHP,Barato2015JSP,Hoppenau2016NJP,Touchette2018PA,Mallmin2021JPAMT,Monthus2021JSMTE}
. For a variety of reasons
detailed below 
and in the accompanying letter \cite{AccompanyingLetter} we here
define $U_\x^h$ with a finite length scale $h>0$, such that $\overline{\rho^U_\x}(t)$
measures the time spent in the region $U_\x^h$ around $\x$ and
$\overline{\f J^U_\x}(t)$ the displacements in the region
$U_\x^h$ around $\x$. Such a definition is in line with that of 
generalized currents in stochastic thermodynamics \cite{Qian_PRE,Horowitz2019NP,Gingrich2016PRL,Gingrich2017JPAMT} except that
we here consider vector-valued currents. Important recent results on
such generalized currents (however, without the notion of coarse graining) may be found in
\cite{Dechant2021PRR,Dechant2018JSMTE,Dechant2018JPAMT,Dechant2020PNASU,Dechant2021PRX}.

\begin{figure}[ht!!]
\centering \includegraphics[width=0.45\textwidth]{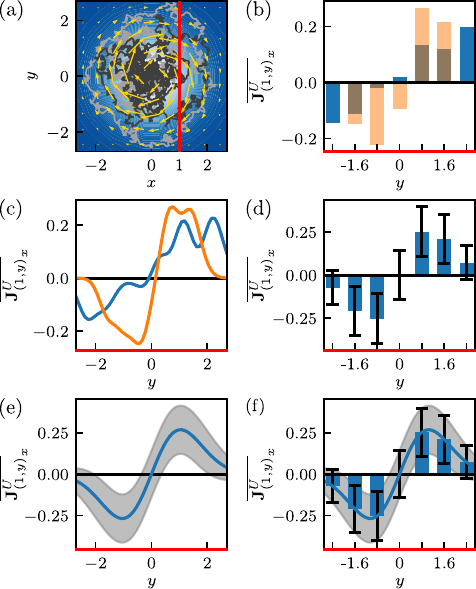}
\caption{(a) Two trajectories (gray) with length $t=5$ in harmonically
  confined rotational flow Eq.~\eqref{OUP} with $\Omega=5$. The steady-state density and current are depicted by the
  color gradient and yellow arrows, respectively. (b) Height of bins
  depict the time-averaged $x$-component of the current with Gaussian
  coarse-graining window Eq.~\eqref{Gauss window} with $h=0.3$
  evaluated for several points on the red line in (a) for the two
  trajectories in (a). This corresponds to time-averaging all local displacements (weighted by $U^h_\x$) within a single trajectory. (c) As in (b) but for the continuum of points on the red line in (a). This can be considered as the $x$-component of the current smoothened over a scale $h$. (d) Mean value $\Es{A}$ and standard deviation $\sqrt{\rm var(A)}$ of $A=\overline{\f J^U_{(1,y)}}_x$ obtained from our result Eq.~\eqref{current component variance result}. This represents the statistics of many histograms as in (b). (e) As in (d) but for continuous $y$ as in (c). (f) Overlaying (d) and (e) shows that the histogram picture is fully contained in the continuous coarse graining procedure.\label{fig:histograms}}
\end{figure}

The fluctuations of $\overline{\rho^U_\x}(t)$ and $\overline{\f J^U_\x}(t)$ may be interpreted as variances of fluctuating histograms. Namely, after ``binning'' into
(hyper)volumes around points $\x$ (or in our language the coarse-graining around $\x$), often carried out on a
grid, each individual trajectory yields a random histogram of occupation fractions or
displacements. That is, the height of bins in the histogram
reflects the time spent or
displacement in said bin accumulated over all visits of the
trajectory until time $t$ for $\overline{\rho^U_\x}(t)$ and $\overline{\f
  J^U_\x}(t)$, respectively, and is a fluctuating quantity due to the
stochasticity of trajectories. The 
variance of these fluctuations quantifies the inference
uncertainty. In Fig.~\ref{fig:histograms} we show such 
histograms inferred from individual trajectories of a two-dimensional harmonically
confined overdamped diffusion in a rotational flow
\begin{align}
  \rmd\x_t=-\begin{bmatrix}1&-\Omega\\\Omega&1\end{bmatrix}\x
  \rmd t+\sqrt{2}\rmd\f W_t,\label{OUP}
\end{align}
with Gaussian window 
\begin{align}
  U^h_\x(\z)=\frac{1}{2\pi h^2}\exp\left[-\frac{(\z-\x)^2}{2h^2}\right].
  \label{Gauss window} 
\end{align}
For this process and window function we analytically solved all
spatial integrals \cite{AccompanyingLetter} entering the results derived below, and numerically evaluated one
remaining time-integral. 

\begin{figure}
  \centering
  \includegraphics[width=0.46\textwidth]{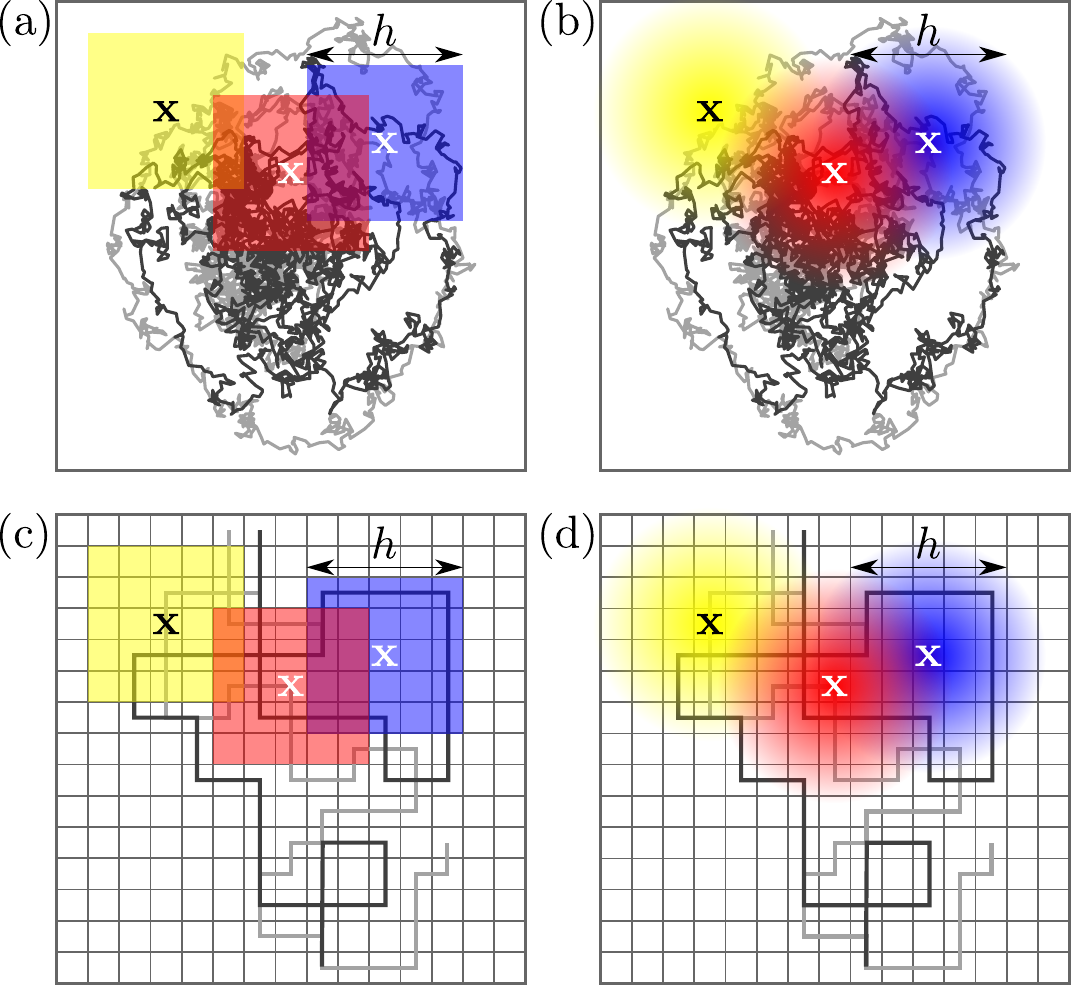}
  \caption{(a) Coarse-graining windows (colors) in the form of an
    indicator function of a rectangle centered at different points
    $\x$ with coarse-graining scale $h$. For each $\x$ and $h$, each
    trajectory (gray lines) gives rise to one value for the (coarse
    grained) time-averaged density and current. Note that the choice
    of $\x$ and $h$ is flexible such that the windows may overlap. (b) Same as (a) but
    with Gaussian coarse-graining windows. (c-d) Coarse-graining windows in the case of trajectory data with a finite
    experimental resolution (grid, gray trajectories). The coarse
    graining scale $h$ should be chosen large compared to the
    resolution to obtain reliable approximations of the
    (coarse-grained) densities and currents. }
      \label{fig:notion}
\end{figure}

The interpretation of the coarse graining captured in or induced by $U^h_\x$ in
Eq.~\eqref{def_current} is flexible; it can represent a projection or
a ``generalized current'' \cite{Qian_PRE,Horowitz2019NP,Gingrich2016PRL,Gingrich2017JPAMT,Dechant2021PRR,Dechant2018JSMTE,Dechant2018JPAMT,Dechant2020PNASU,Dechant2021PRX} or may
be thought of as a spatial
smoothing of the empirical current and density as shown in Fig.~\ref{fig:histograms}c,e and
Fig.~\ref{fig:notion}, also for the case of 
a finite experimental resolution. Our main focus
here is the smoothing aspect in the context of uncertainty
of $\ps(\x),\js(\x)$ and steady-state dissipation from individual trajectories. 
Note that some form of coarse graining or smoothing
is in fact required in order for the quantities in 
Eq.~\eqref{def_current} to be well defined \cite{AccompanyingLetter}.  
A suitable smoothing decreases the uncertainty of the estimate and, if
varied over sufficiently many $h$ and $\x$ (see also
Fig.~\ref{fig:histograms}c,e) instead of simply "binning",
one does not necessarily lose information (as compared to input data).
Moreover, a
systematic variation of the scale $h$ may reveal more information
about $\overline{\rho^U_\x}(t)$ and $\overline{\f J^U_\x}(t)$. The
same reasoning is found to apply to generalized thermodynamic currents
and allows for an improved inference of dissipation, see \cite{AccompanyingLetter} and 
below.

The present work is an extended expos\'e of the conceptual
and
technical 
background that is required to understand and materialize
the above observations. It accompanies the letter
\cite{AccompanyingLetter} but does not duplicate any information. Several additional explanations, illustrations and applications are given here.

The article is structured as follows. In Sec.~\ref{Theory} we lay
out the theoretical background on stochastic differential equations in
the It\^o, Stratonovich and anti-It\^o interpretations and the
corresponding equations for the probability densities. We furthermore
decompose the drift and steady-state current into conservative and
non-conservative (i.e.\ irreversible) contributions and introduce
dissipation. In Sec.~\ref{Dual} we prove a generalize time-reversal symmetry called
``dual-reversal symmetry''. In Sec.~\ref{derivation} we derive our main results for the steady-state
(co)variances of $\overline{\rho^U_\x}(t)$ and $\overline{\f
  J^U_\x}(t)$ and interpret them in terms of initial- and end-point
currents and increments. \blue{We then use these results to explicitly evaluate the limit $h\to 0$ of no coarse graining in Sec.~\ref{sec:h to 0}, where we find that fluctuations diverge in $d\ge2$-dimensional space.} In Sec.~\ref{APPLI} we use current fluctuations to infer
steady-state dissipation via the Thermodynamic Uncertainty Relation (TUR) \cite{Barato2015PRL,Dechant2018JSMTE} with an emphasis on the importance of the coarse-graining
scale $h$. In particular we demonstrate and explain the existence of a
thermodynamically optimal coarse graining. In
Sec.~\ref{sec:symmetry} we discuss symmetries obeyed by the
(co)variances and explain how the results simplify in thermodynamic
equilibrium\blue{, and in Sec.~\ref{cont eq} we present a continuity
  equation for coarse grained empirical densities and currents}. In Sec.~\ref{short and long times} we present asymptotic results for short and long trajectories and
give results for \finalchange{the central-limit regime}. We conclude \blue{with an
  outlook beyond overdamped dynamics in Sec.~\ref{sec:outlook} by
  considering underdamped systems as well as experimental data derived
  from particle-tracking experiments in biological cells, and}
with a summary and perspectives for the future.

\section{Theory}\label{Theory}

\subsection{Set-up -- overdamped Langevin dynamics}
In this section we provide background on the equations of motion for
the coordinate $\x_\tau$ highlighting the differences between the It\^o, Stratonovich, and
anti-It\^o interpretations, and for their corresponding conditional probability density
functions of a transition $\x_0\to \x$. 

We consider time-homogeneous (i.e.\ coefficients do not explicitly
depend on time) overdamped Langevin dynamics in $d$-dimensional space
with (possibly) multiplicative noise
\cite{Gardiner1985,Pavliotis2014TiAM} described by the
thermodynamically consistent \cite{Pigolotti2017PRL,Hartich2021PRX}
anti-It\^o~(or~H\"anggi-Klimontovich~\cite{anti,anti2})
stochastic differential equation
\begin{align}
\rmd\x_\tau=\f F(\x_\tau)\rmd\tau+\bsig(\x_\tau)\circledast\rmd\f W_\tau,
\label{SDE} 
\end{align}
where $\rmd\f W_\tau$ is the increment of a $d$-dimensional Wiener
processes (i.e.\ white noise) with zero mean and covariance
$\langle\rmd W_{\tau,i}\rmd
W_{\tau',j}\rangle=\delta(\tau-\tau')\delta_{ij}d\tau$. The noise
amplitude is related to the diffusion coefficient via $\f D(\x)\equiv \bsig(\x)\bsig(\x)^T/2$.
We assume the drift field $\f F(\x)$ to be  smooth and sufficiently
confining, such that the anti-It\^o (end-point)
 convention $\circledast\rmd\f W_\tau=\f W_\tau-\f W_{\tau-d\tau}$
 guarantees the existence of a steady-state probability 
 density $\ps(\x)=\rme^{-\phi(\x)}$ and steady-state current $\js(\x)$, and yields \blue{the thermodynamically consistent}
Boltzmann-Gibbs (equilibrium) statistics when $\f D(\x)^{-1}\f F(\x)=-\nabla \phi(\x)$ is a
potential force. 

The anti-It\^o equation \eqref{SDE} can equivalently be rewritten as an It\^o equation with an adapted drift as,
\begin{align}
\rmd\x_t&=\f F(\x_t)\rmd t+\bsig(\x_t)\circledast\rmd\f W_t\nonumber\\
&=\f F(\x_t)\rmd t+\left[\left\{\nabla^T\sqrt{2\f D(\x_t)}\right\}\rmd\x_t\right]\cdot\rmd\f W_t\nonumber\\
&\quad+\sqrt{2\f D(\x_t)}\rmd\f W_t\nonumber\\
&=\left[\f F(\x_t)+\left\{\nabla^T\f D\right\}(\x_t)\right]\rmd
t+\sqrt{2\f D(\x_t)}\rmd\f W_t,
\label{SDE ito} 
\end{align}
where the brackets $\{\cdot\}$ throughout denote that the differential
operator only acts within the bracket and $\sqrt{2\f D(\x_t)}$ represents the matrix $\bsig(\x_t)$.
At this point several remarks are in order. First, the anti-It\^o  interpretation
of the stochastic differential equation \eqref{SDE} as well as the Stratonovich integral in
Eq.~\eqref{def_current} are both required
for thermodynamic consistency. Second, there is no
difference between the interpretations of  Eq.~\eqref{SDE} if $\f
D(\x)=\f D$ is a constant matrix, i.e.\ the convention only matters
for multiplicative noise. However, even in this case the Stratonovich integral in
Eq.~\eqref{def_current} is required for thermodynamic consistency of
the empirical current and to use it as an estimator of $\js(\x)$.


The Fokker-Planck equation for the conditional probability density
$G(\x,t|\y)$ to be at a point $\x$ at time $t$ after starting at $\y$
that corresponds to 
Eqs.~\eqref{SDE} and \eqref{SDE ito} reads 
\begin{align}
\partial_tG(\x,t|\y)&=[-\nabla_\x\cdot\f F(\x)+\nabla_\x^T\f
  D(\x)\nabla_{\x}]G(\x,t|\y)\nonumber\\
&\equiv L(\x)G(\x,t|\y),
\end{align}
which satisfies a continuity equation 
$(\partial_t+\nabla_{\x}\cdot\bj_{\x})G(\x,t|\y)=0$, where
\begin{equation}
\bj_\x\equiv \f F(\x)-\f D(\x)\nabla_{\x}.
\end{equation}
\blue{Decomposing of the drift $\f F(\x)$ into reversible $\f F^{\rm rev}(\x)=-\f D(\x)\{\nabla\phi\}(\x)$ and irreversible $\f F^{\rm irrev}(\x)=\f F(\x)-\f F^{\rm rev}(\x)$ parts} translates to a decomposition of
$\bj_\x$ into a gradient part $\bj^g(\x)$ and steady-state-current contributions, namely $\bj_\x=\f F^{\rm irrev}(\x)+\f F^{\rm rev}(\x)-\f D(\x)\nabla_\x$. This is rewritten using
\begin{align}
\bj^g(\x)
&\equiv\f F^{\rm rev}(\x)-\f D(\x)\nabla\nonumber\\
&=\f D(\x)\left\{\nabla\log(\ps(\x)\right\}-\f D(\x)\nabla\nonumber\\
&=\f D(\x)\ps^{-1}(\x)\left\{\nabla\ps(\x)\right\}-\f D(\x)\nabla\nonumber\\
&=-\f D(\x)\left[\ps(\x)\{\nabla\ps^{-1}(\x)\}-\nabla\right]\nonumber\\
&=-\f D(\x)\ps(\x)\nabla\ps^{-1}(\x),
\end{align}
where we have used that
${\{\nabla\ps(\x)^{-1}\}}{=}{-\ps^{-2}(\x)\{\nabla\ps\}(\x)}$ implies
${\{\nabla\ps\}(\x)}{=}{-\ps^2(\x)\{\nabla\ps^{-1}\}(\x)}$).\ \blue{Therefore
we have 
$\bj^g(\x)\ps(\x)=0$, such that the definition of the steady-state
current $\js(\x)\equiv \bj(\x)\ps(\x)$ with $\bj_\x=\bj^g(\x)+\f
F^{\rm irrev}(\x)$ implies $\f F^{\rm irrev}(\x)=\ps^{-1}(\x)\js(\x)$ and we obtain
\begin{align}
\bj_\x
&=\bj^g_\x+\ps^{-1}(\x)\js(\x)\nonumber\\
&=-\ps(\x)\f D(\x)\nabla_\x\ps^{-1}(\x)+\ps^{-1}(\x)\js(\x).
\label{current_operator_decomposed}
\end{align}
}Moreover, note that the steady-state two-point density $P_\y(\x,t)\equiv G(\x,t|\y)\ps(\y)$
also satisfies the same Fokker-Planck equation as $G(\x,t|\y)$.

Finally, if the process is irreversible, i.e.\ $\f F^{\rm irrev}(\x)\ne
\f 0$ the steady state is dissipative with an average total entropy
production rate $\dot\Sigma$ given by \cite{Spinney2012PRE,Seifert2012RPP}
\begin{align}
\dot\Sigma&=\int\rmd\x\f F^{\rm irrev}(\x)\cdot\f D^{-1}(\x)\f F^{\rm
  irrev}(\x)\ps(\x)\nonumber\\
&=\int\rmd\x\frac{\js^T(\x)}{\ps(\x)}\f D^{-1}(\x)\js(\x),
\label{EPR} 
\end{align}
which can be obtained as the mean value of a sum over steady-state expectations of the respective $i$-th component of $\overline{\f J^{U_i}_\x}(t)$ in Eq.~\eqref{def_current} with $U_i=(\f F^{\rm irrev}(\x)^T\f D^{-1}(\x))_i$.


Note that by adopting the It\^o or Stratonovich conventions instead of the anti-It\^o convention in Eq.~\eqref{SDE} one obtains a different Fokker-Planck
equation with a different steady-state density. In particular,
$L^{\rm Ito}(\x)=-\nabla_\x\cdot\f
F(\x)+\sum_{i,j=1}^d\partial_i\partial_jD_{ij}(\x)$ 
and
$L^{\rm Strato}(\x)=L(\x)/2+L^{\rm Ito}(\x)/2=-\nabla_\x\cdot\f
F(\x)+\sum_{i,j=1}^d\partial_i\sqrt{D_{ij}(\x)}\partial_j\sqrt{D_{ij}(\x)}$
and the respective steady-state densities $\ps^{\rm Ito}(\x)$ and
$\ps^{\rm Strato}(\x)$ depend explicitly on $\f D(\x)$ and are
therefore in general not thermodynamically consistent since the steady state deviates from Gibbs-Boltzmann statistics (e.g.\ in dimension one we have \blue{$\ps^{\rm Ito}(x)\propto\ps^\textrm{anti-Ito}(x)/D(x)$
and $\ps^{\rm Strato}(x)\propto\ps^\textrm{anti-Ito}(x)/\sqrt{D(x)}$,
respectively, where the deviation from $\ps^\textrm{anti-Ito}(x)$} cannot be absorbed in the normalization if $D(x)$ depends on $x$).

\section{Generalized time-reversal symmetry}\label{Dual}
It will later prove useful to take into account a form of generalized time-reversal
symmetry obeyed by Eq.~\eqref{SDE} called ``continuous time reversal'' or ``dual-reversal symmetry''
\cite{Hatano2001PRL,Dechant2021PRR}. Analogous generalized symmetries were also
found in deterministic systems (see e.g.\ \cite{RondoniTimeReversal}).
Generalized time-reversal symmetry relates forward dynamics in
non-equilibrium steady states to time-reversed dynamics in an ensemble
with inverted irreversible steady-state current, i.e.\ in an ensemble
with $\f F^{\rm irrev}\to-\f F^{\rm irrev}$ or equivalently
$\js\to-\js$. The dual-reversal symmetry for the two-point probability densities states that
\begin{equation}
  G(\x,t|\y)\ps(\y)=G^{-\js}(\y,t|\x)\ps(\x),
  \label{dual_rev}
\end{equation}
or equivalently $G^{-\js}(\x,t|\y)\ps(\y)=G(\y,t|\x)\ps(\x)$ where
$G^{-\js}(\y,t|\x)$ is the conditional probability density of the
process with drift $\f F^{-\js}(\x)\equiv\f F^{\rm rev}(\x)-\f F^{\rm irrev}(\x)$ instead of $\f F(\x)=\f F^{\rm rev}(\x)+\f F^{\rm irrev}(\x)$.  At equilibrium, i.e.\ $\js(\x)=\f 0$ (for all $\x$),
this symmetry simplifies to the well known time-reversal symmetry
called ``detailed balance'' condition for two-point densities.
%
We here provide an original and intuitive proof of Eq.~\eqref{dual_rev} that proceeds entirely in
continuous space and time, based on the decomposition of currents
Eq.~\eqref{current_operator_decomposed}. The Fokker-Planck operator
$L(\x)=-\nabla_\x\cdot\bj_\x$, using the decomposition Eq.~\eqref{current_operator_decomposed} and multiplying by $\ps$ from the right side, reads
\begin{align}
L(\x)\ps(\x)=-\nabla_\x\cdot\js(\x)+\nabla_\x^T\ps(\x)\f D(\x)\nabla_\x.
\end{align}
Taking the adjoint gives (since $\f D=\f D^T$)
\begin{align}
\ps(\x)L^\dag(\x)&=[L(\x)\ps(\x)]^\dag\nonumber\\
&=\js(\x)\cdot\nabla_\x+\nabla_\x^T\ps(\x)\f D(\x)\nabla_\x.
\end{align}
Since for the steady state density $L\ps=0$, $\js$ is divergence free $\{\nabla_\x\cdot\js(\x)\}=0$ and we have $\nabla_\x\cdot\js(\x)=\js(\x)\cdot\nabla_\x$. Thus we see the symmetry under inversion $\js\rightarrow-\js$
\begin{align}
\ps(\x)L^\dag(\x)=L^{-\js}(\x)\ps(\x).
\label{dual reversal FPO} 
\end{align}
Under detailed balance $\js=\f 0$, i.e.\ $L^{-\js}=L$, and
$\ps(\x)L^\dag(\x)=L(\x)\ps(\x)$ which implies the time-reversal
symmetry $G(\x,t|\y)\ps(\y)=G(\y,t|\x)\ps(\x)$
\cite{Risken1996,Pavliotis2014TiAM,Durrett_Stoch}.
Eq. \eqref{dual reversal FPO} implies for all integers $n\ge 1$ that
$\ps(\x)[L^\dag(\x)]^n=[L(\x)^{-\js}]^n(\x)\ps(\x)$, and consequently
for all $t\ge 0$ that $\ps(\x)\exp[L^\dag(\x)
  t]=\exp[L^{-\js}(\x)t]\ps(\x)$. Applying this operator equation to
the initial condition $\delta(\y-\x)$ and using
$\ps(\x)\delta(\y-\x)=\ps(\y)\delta(\y-\x)$ as well as that $L^\dag$
propagates the initial condition as $G(\y,t|\x)=\exp[L^\dag(\x)
  t]\delta(\y-\x)$ while $L^{-\js}$ propagates the final point in the
ensemble with $\js$ inverted $G^{-\js}(\x,t|\y)=\exp[L^{-\js}(\x)
  t]\delta(\y-\x)$, we obtain the dual reversal symmetry in Eq.~\eqref{dual_rev}.
This generalized time-reversal symmetry relates the dynamics in the time-reversed
ensemble to the propagation in the ensemble with reversed current, or
equivalently, the forward dynamics to the propagation with concurrent time and
$\js$-reversal. While at equilibrium (i.e.\ under detailed balance,
$\js=\f 0$) the forward dynamics is indistinguishable from the
time-reversed dynamics, the statement Eq.~\eqref{dual_rev} (if
generalized to all paths (see e.g.\ \cite{Dechant2021PRR}) means that forward
dynamics (with $\js$) is indistinguishable from
backwards/time-reversed dynamics with reversed $\js\to-\js$
(i.e.\ $\js(\x)\to-\js(\x)$ at all $\x$). We will later use
this dual-reversal symmetry to understand the fluctuations of observables
that involve (time-integrated) currents in non-equilibrium steady states.  


\section{Derivation of the main results, initial- and final-point currents and their application to density-current correlations}\label{derivation}
\subsection{Mean empirical density and current}
Although the time-averaged density and current defined in
Eq.~\eqref{def_current} are functionals with complicated statistics,
their mean values can be readily computed. Throughout the paper we
will assume steady-state initial conditions, i.e.\ initial conditions
drawn from $\ps(\x')$, denoted by $\Es{\cdot}$. This renders mean
values time-independent and we have (see also \cite{Maes2008EEL})
\begin{align}
  \Es{\overline{\rho^U_\x}(t)}&=\frac1t\int_0^t\rmd\tau\Es{U^h_\x(\x_\tau)}\nonumber\\  
&=\frac{1}{t}\int_0^t\rmd\tau\int\rmd\z U^h_\x(\z)\ps(\z)\nonumber\\
&=\int\rmd\z U^h_\x(\z)\ps(\z),\label{mean_rho}
\end{align}
and by rewriting the Stratonovich-integration
$\circ\rmd\x_\tau$ 
in terms of It\^o
integration as $U^h_\x(\x_\tau)\circ\rmd\x_\tau=U^h_\x(\x_\tau)\rmd\x_\tau+\frac{1}{2}\rmd U^h_\x(\x_\tau)\rmd\x_\tau$, 
where $\rmd\x_\tau\rmd\x_\tau^T/2=\f D(\x_\tau)\rmd\tau$ and thus $\rmd U^h_\x(\x_\tau)\rmd\x_\tau/2=\f D(\x_\tau)\{\nabla U^h_\x\}(\x_\tau)\rmd\tau$,
\begin{align}
&\Es{\overline{\f J^U_\x}(t)}=\frac{1}{t}\int_0^t\Es{U^h_\x(\x_\tau)\circ\rmd\x_\tau}\nonumber\\
&=\frac{1}{t}\int_{\tau=0}^{\tau=t}\Es{U^h_\x(\x_\tau)\rmd\x_\tau}+\frac{1}{t}\int_{\tau=0}^{\tau=t}\frac{1}{2}\Es{\rmd U^h_\x(\x_\tau)\rmd\x_\tau}\nonumber\\
&=\frac{1}{t}\int_0^t\rmd\tau\int\rmd\z \ps(\z)\big[U^h_\x(\z)\f F(\z)+\left\{\nabla^T_\z\f D(\z)\right\}U^h_\x(\z)\nonumber\\
&+\f D(\z)\left\{\nabla_\z U^h_\x(\z)\right\}\big]+\frac{1}{t}\int_{\tau=0}^{\tau=t}\Es{U^h_\x(\x_\tau)\sqrt{2\f D(\x_\tau)}\rmd\f W_\tau}.\label{mean_j_part1}
\end{align}
Note that the mean value involving $\rmd\f W_\tau$ vanishes since this
It\^o-noise increment has zero mean and is uncorrelated with functions
of $\x_\tau$, i.e. $\langle f(\x_\tau)\rmd\f W_\tau\rangle=\langle
f(\x_\tau)\rangle\langle\rmd\f W_\tau\rangle=0$. Integrating by parts and
using that $\f D(\z)=\f D^T(\z)$ is symmetric we get
\begin{align}
\Es{\overline{\f J^U_\x}(t)}&=\int\rmd\z \ps(\z)\left[U^h_\x(\z)\f F(\z)+\nabla^T_\z\f D(\z)U^h_\x(\z)\right]\nonumber\\
&=\int\rmd\z U^h_\x(\z)\left[\f F(\z)-\f D(\z)\nabla_{\z}\right]\ps(\z)\nonumber\\
&=\int\rmd\z U^h_\x(\z)\bj_\z\ps(\z)=\int\rmd\z U^h_\x(\z)\js(\z).
\label{mean_j_part2}
\end{align}
Note that if we had defined Eq.~\eqref{def_current} with an It\^o
integral instead of the Stratonovich, we would miss the $\f
D(\z)\nabla_\z$-term and would \emph{not} get $\bj_\z$ and thus
$\js$, not even for additive noise. The Stratonovich integral is therefore required for consistency. 

The interpretation of the steady-state mean values in
Eqs.~\eqref{mean_rho} and \eqref{mean_j_part2} is immediate --- the mean time-averaged density and current are (at least for positive normalized windows) the steady-state density $\ps$ and current $\js$ averaged over the coarse-graining window function $U^h_\x$.

\subsection{(Co)variances of empirical density and current}
Since fluctuations
\cite{Barato2015PRL,Dechant2018JSMTE,Burov2011PCCP,Lapolla2020PRR,Qian_PRE,Horowitz2019NP,Gingrich2016PRL,Gingrich2017JPAMT,Dechant2021PRR}
(and correlations \cite{Dechant2021PRX}) 
play a crucial role in time-average statistical mechanics and
stochastic thermodynamics, we discuss (co)variances of coarse-grained time-averaged densities and currents (recall
the interpretation of the variance within the ``fluctuating
histogram'' picture in Fig.~\ref{fig:histograms}).

To keep the notation tractable we introduce the integral operator
\begin{align}
\!\integrals^{t,U}_{\x\y}[\,\cdot\,]\equiv\frac{1}{t^2}\!\int_0^t\!\rmd t_1\!\!\int_{t_1}^t\!\rmd t_2\!\!\int\!\rmd\z U^h_\x(\z)\!\!\int\!\rmd\z' U^h_\y(\z')[\,\cdot\,],
\label{int_op}
\end{align}
with the convention $\int_{t_1}^t\rmd t_2\delta(t_2-t_1)=1/2$. Note
that other conventions would only change the appearance of intermediate steps but not the final result. We define the two-point steady-state covariance according to
\cite{AccompanyingLetter} as
\begin{equation}
C^{\x\y}_{AB}(t)\equiv\langle A_\x(t)B_\y(t)\rangle_{\rm s} - \langle
A_\x(t)\rangle_{\rm s}\langle B_\y(t)\rangle_{\rm s}\, ,
\label{covar_g}
\end{equation}
where $A$ and $B$ are henceforth either $\overline{\rho^U}$ or
$\overline{\f J^U}$, respectively. We refer to the case when $A\ne B$
or $\x\ne\y$ as (linear) ``correlations'' and to the case $A=B$ with
$\x=\y$ as ``fluctuations'' whereby we adopt the convention
${\mathrm{var}^{\x}_{A}(t)\equiv C^{\x\x}_{AA}(t)}$.  Note that for
simplicity and enhanced readability 
we only assume coarse-graining windows $U^h_\x$ and $U^h_\y$ where \blue{the shape is fixed but the center points $\x,\y$ may differ}. All results equivalently
hold for window functions whose shape and $h$ differs as well. 

We now address correlations $C^{\x\y}_{\rho\rho}$ of the
coarse-grained time-averaged density at points $\x$ and $\y$, which
corresponds to the density variance when $\x=\y$. To do so, first
consider the (mixed) second moment
\begin{align}
\Es{\overline{\rho^U_\x}(t)\overline{\rho^U_\y}(t)}&=\int_0^t\rmd\tau\int_0^t\rmd\tau'\Es{U^h_\x(\x_\tau)U^h_\y(\x_\tau')}.
\end{align}
The expectation value corresponds to an integration over the two-point
probability density to have $\x_{\tau}=\z$ and $\x_{\tau'}=\z'$ given
by the two-point function $P_\z(\z',\tau'-\tau)\equiv G(\z',\tau'-\tau|\z)\ps(\z)$ for $\tau'>\tau$ and
$P_{\z'}(\z,\tau-\tau')$ for $\tau'<\tau$.  We relabel the times
$\tau,\tau'$ as $t_1<t_2$ and use the integral operator in Eq.~\eqref{int_op} to obtain
\begin{align}
\Es{\overline{\rho^U_\x}(t)\overline{\rho^U_\y}(t)}&=\integrals^{t,U}_{\x\y}\left[P_\z(\z',t_2-t_1)+P_\z(\z',t_2-t_1)\right].
\end{align}
Since the argument only depends on time differences $t'=t_2-t_1\ge 0$
the integral operator Eq.~\eqref{int_op} simplifies to
\begin{align}
\!\!\integrals^{t,U}_{\x\y}[\,\cdot\,]\equiv\frac{1}{t}\!\int_0^t\!\!\!\rmd t'\!\left(1-\frac{t'}{t}\right)\!\!\int\!\rmd\z U^h_\x(\z)\!\!\int\!\!\rmd\z' U^h_\y(\z')[\,\cdot\,].
\label{int_op_simpl}
\end{align}
To obtain the correlation we subtract the mean values (see Eq.~\eqref{mean_rho}) which (noting that $(1/t)\int_0^t\rmd t'(1-t'/t)=1/2$) gives 
\begin{align}
  \!\!C^{\x\y}_{\rho\rho}(t)=\integrals^{t,U}_{\x\y}\left[P_\z(\z',t')+P_{\z'}(\z,t')-2\ps(\z)\ps(\z')\right],
  \label{density-density result} 
\end{align}
which has been derived before
\cite{Darling1957TAMS,Lapolla2020PRR}. Eq.~\eqref{density-density
  result} simplifies further for $\x=\y$ as well as under detailed
balance and is also symmetric under $\js\to-\js$, all of which will be discussed in Sec.~\ref{sec:symmetry}.

The interpretation of Eq.~\eqref{density-density result} (see also
\cite{Lapolla2020PRR}) is that all paths from $\z$ to $\z'$
(i.e.\ from $U^h_\x$ to $U^h_\y$) and vice versa from $\z'$ to $\z$,
in time $t'=t_2-t_1$ contribute according to their correlation to
$C^{\x\y}_{\rho\rho}(t)$.  These contributions are integrated over all
possible time differences and pairs of points within $U^h_\x$ and
$U^h_\y$, respectively.

\begin{figure}[ht!!]
  \centering
  \includegraphics[width=.45\textwidth]{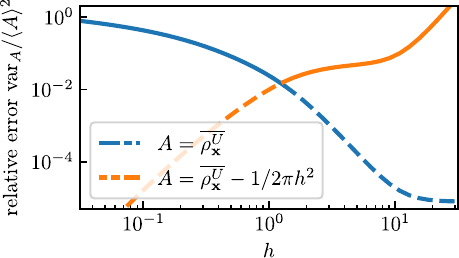}
  \caption{Relative error of $\overline{\rho^U_\x}(t)$ (blue line) compared to the relative error of $[\overline{\rho^U_\x}(t)-U^h_\x(\x)]$ (orange line) as a function of the coarsening scale $h$ for the rotational flow Eq.~\eqref{OUP} with $\Omega=3$ for time $t=10$ with a Gaussian window function Eq.~\eqref{Gauss window} around $\x=(1,0)^T$ with width $h$, i.e.\ $U^h_{\x}(\x)=(2\pi h^2)^{-1}$.
The intersection point of blue and orange lines at $h\approx 1.3$ yields an
"optimal coarse graining" where the maximum of the two lines (solid line) is
minimal, whereas the maximum of the relative errors diverges as $h\to\infty$ since $\Es{A}\to 0$ and diverges logarithmically for $h\to 0$ \blue{(as we will see in Eq.~\eqref{bounds h to 0})}.
\label{fig:rel_err_rho}}
\end{figure}

We now explore the important effect of coarse graining over the windows
$U^h_\x$ for the inference of $\ps(\x)$ from noisy individual
trajectories.  If one wants to reliably infer the (coarse-grained) steady-state density from $\overline{\rho^U_\x}(t)$ the relative error  ${\rm var}_\rho/\langle \overline{\rho^U_\x}(t)\rangle^2$ should be small. We
have shown that $\lim_{h\to 0}{\rm
  var}_\rho/\langle \overline{\rho^U_\x}(t)\rangle^2=\infty$
\cite{AccompanyingLetter} and Fig.~\ref{fig:rel_err_rho} (blue line) demonstrates
that ${\rm
  var}_\rho/\langle \overline{\rho^U_\x}(t)\rangle^2$ decreases with
increasing $h$. However, such a decrease does not guarantee an improved
inference. Namely, as $h\to\infty$ the time to spent in the region
around $\x$ tends to $t$ and $U^h_\x$ becomes constant on a large region and hence $\overline{\rho^U_\x}(t)\to
U^h_\x(\x)$ which contains no information about $\ps(\x)$. Therefore, to reliably infer that $\overline{\rho^U_\x}$
significantly deviates from $U^h_\x(\x)$ we must also consider the
relative error of $[\overline{\rho^U_\x}-U^h_\x(\x)]$ depicted in
Fig.~\ref{fig:rel_err_rho} (orange line). There exists an
"optimal coarse graining" where the uncertainty of simultaneously
inferring $\overline{\rho^U_\x}$ and $\overline{\rho^U_\x}-U^h_\x(\x)$ is
minimal (minimum of the solid lines in Fig.~\ref{fig:rel_err_rho}) which represents the most reliable and informative estimate of $\overline{\rho^U_\x}$. In Sec.~\ref{APPLI} we will turn to an analogous ``optimal
coarse graining'' with respect to current variances and a system's dissipation.

We now consider coarse-grained time-averaged currents. To compute the correlation of the current at a point $\x$ and the density at $\y$ we need to consider 
\begin{align}
\Es{\overline{\f J^U_\x}(t)\overline{\rho^U_\y}(t)}&=\int_0^t\rmd\tau\int_{\tau'=0}^{\tau'=t}\Es{U^h_\x(\x_\tau)U^h_\y(\x_\tau')\circ\rmd\x_{\tau'}}.
\end{align}
Relabeling with $t_1\le t_2$, introducing the notation
\begin{align}
\E{\cdots}^{\x_{t_1}=\z}_{\x_{t_2}=\z'}\equiv\Es{\delta(\x_{t_1}-\z)\delta(\x_{t_2}-\z')\cdots},\label{notation_end-points} 
\end{align}
and considering the Stratonovich increments
\begin{align}
  \circ\rmd\x_\tau\equiv\x_{\tau+\rmd\tau/2}-\x_{\tau-\rmd\tau/2},
  \label{notation_strato-increment} 
\end{align}
and subtracting the mean values \eqref{mean_rho} and \eqref{mean_j_part2}, we can write the correlation as
\begin{align}
\f C^{\x\y}_{\f
  J\rho}(t)=\integrals^{t,U}_{\x\y}&\Bigg[\frac{{\E{\circ\rmd\x_{t_1}}}^{\x_{t_2}=\z'}_{\x_{t_1}=\z}}{\rmd
    t_1}\nonumber\\
  & +\frac{{\E{\circ\rmd\x_{t_2}}}^{\x_{t_2}=\z}_{\x_{t_1}=\z'}}{\rmd    t_2}- 2\js(\z)\ps(\z')\Bigg].
\label{correlation ansatz} 
\end{align}
Eq.~\eqref{correlation ansatz} is harder to compute and more difficult to interpret as compared to
$C^{\x\y}_{\rho\rho}(t)$ (see Eq.~\eqref{density-density result}). The quantities involving Stratonovich increments
\blue{characterize the mean initial- and final
displacements of ``pinned'' paths of duration $t_2-t_1$ conditioned on the
initial and final points $\z,\z'$ or $\z',\z$, respectively. Note that }$\z$ always denotes the
point where the increment occurs.  Via the integral operator in
Eq.~\eqref{int_op} or \eqref{int_op_simpl}  the $\z$ variable is
integrated over $U^h_\x(\z)$, i.e.\ in $\f C^{\x\y}_{\f J\rho}(t)$ the
variable $\z$ corresponds to the \blue{window at $\x$ where} the (coarse-grained)
current is evaluated. \blue{Therefore, correlations between a current and a density depend on integrals over conditioned initial-point increments at a point $\z$ at time $t_1$, and conditioned final-point increments, also at $\z$, at time $t_2>t_1$. We define the increments divided by $\rmd t_i$ to be the "initial- and final-point currents",}
\begin{align}
\f j_{\rm in}(\z',t_2-t_1;\z)&\equiv\frac{\E{\circ\rmd\x_{t_1}}^{\x_{t_2}=\z'}_{\x_{t_1}=\z}}{\rmd t_1}\nonumber\\
\f j_{\rm fi}(\z,t_2-t_1;\z')&\equiv\frac{\E{\circ\rmd\x_{t_2}}^{\x_{t_2}=\z}_{\x_{t_1}=\z'}}{\rmd t_2}.
\label{IPC_FPC}
\end{align}
In order to understand the correlation in Eq.~\eqref{correlation
  ansatz} we must therefore understand initial- and final-point
currents.  This is \emph{a priori} not easy, since initial-point
currents involve both, spatial increments at $t_1$ and probabilities of reaching a
final point at time $t_2>t_1$, which involves non-trivial correlations
--- a given displacement affects (and thus correlates with) the
probability to reach the final point. We will derive a statement
(``Lemma'') in the next subsection that solves all mathematical
difficulties related to this issue\blue{, without resorting to Feynman-Kac and path-integral methods as in Ref.~\cite{ArxivJPhysA}}. Then we will make intuitive sense
of the result by exploiting the dual-reversal symmetry in
Eq.~\eqref{dual_rev}. 
 
Before doing so, we also consider the scalar current-current
covariance $C^{\x\y}_{\f J\cdot\f J}(t)$ (note that the complete fluctuations and correlations
of $\overline{\f J^U_\x}(t)$ are characterized by the $d\times d$
covariance matrix with elements $(\mathsf{C}^{\x\y}_{\f J \f
  J}(t))_{ik}=C^{\x\y}_{{\rm J}_i  {\rm J}_k}(t)$; here we focus on
the scalar case $C^{\x\y}_{\f J\cdot \f J}(t)\equiv
\mathrm{Tr}\mathsf{C}^{\x\y}_{\f J \f J}(t)$). Notably, almost all
results remain completely equivalent for other elements of the
covariance matrix, scalar products simply have a slightly more intuitive geometrical interpretation and notation.
Writing down the definition and using the notations as in the steps towards Eq.~\eqref{correlation ansatz} we immediately arrive at
\begin{align}
C^{\x\y}_{\f J\cdot\f J}(t)=&
\integrals^{t,U}_{\x\y}\Bigg[\frac{{\E{\circ\rmd\x_{t_1}\cdot\circ\rmd\x_{t_2}}}^{\x_{t_2}=\z'}_{\x_{t_1}=\z}}{\rmd
    t_1\rmd t_2}\nonumber\\
  &+\frac{{\E{\circ\rmd\x_{t_1}\cdot\circ\rmd\x_{t_2}}}^{\x_{t_2}=\z}_{\x_{t_1}=\z'}}{\rmd
    t_1\rmd t_2}-2\js(\z)\cdot\js(\z')\Bigg],
\label{current-current ansatz} 
\end{align}
which is similar to the correlation in Eq.~\eqref{correlation ansatz}
but involves an average over scalar products of initial- and
final-point increments along individual trajectories ``pinned'' at
initial- and end-points. We will return to Eq.~\eqref{current-current
  ansatz} and solve for these increments in Subsec.~\ref{subsec:current-current} upon having explained the
density-current correlation. 

\subsection{Lemma}\label{subsec:lemma} 
To be able to treat expressions involving the increments correlated
with future positions, we need a technical lemma that will turn out to
be very powerful and central to all calculations. \blue{The required
  statement can also be obtained from the more general concept of Doob
  conditioning
  \cite{Doob,Chetrite2014AHP,Pigolotti2017PRL,Dechant2021PRR}, but
  here we provide a direct proof.} Consider an It\^o
noise increment 
$\sqrt{2\f D(\x_\tau)}\rmd\f W_\tau$  (or equivalently
$\bsig(\x_\tau)\rmd\f W_\tau$) with $\rmd\f W_\tau=\f
W_{\tau+\rmd\tau}-\f W_\tau$. In the following we will need to
compute the expected values involving expressions like
\begin{align}
\star=\Es{\left[\sqrt{2\f D(\x_\tau)}\rmd\f W_\tau\right]_k U(\x_\tau)V(\x_{\tau'})},\label{expectation term} 
\end{align}
where $U(\x')$ and $V(\x')$ are arbitrary differentiable, square
integrable functions, the subscript $k$ denotes the $k$-th component,
and the subscript ${\rm s}$ denotes that the process evolves from
$\ps(\x')$. Correlations of $\rmd\f W_\tau=\f W_{\tau+\rmd\tau}-\f
W_\tau$ with any function of $\x_{\tau'}$ at a time $\tau'\le\tau$
vanish by construction of the Wiener process (it has nominally
independent increments).  However, correlations with functions at
$\tau'>\tau$ are nontrivial. 

Note that given an initial point $\f x_0=\z$ and setting $\sqrt{2\f
  D(\z)}\rmd\f W_{\f 0}=\beps$, the It\^o/Langevin Eq.~\eqref{SDE ito}
predicts a displacement $\rmd\x_0(\z,\beps)=[\f F(\z)+\nabla_\z^T\f
  D(\z)]\rmd t'+\beps$. With this we can write the expectation in Eq.~\eqref{expectation term} for $\tau=0<t'=\tau'$ as $\varepsilon_k$ integrated over the probability to be at points $\z,\z+\rmd\x_0(\z,\beps),\z'$ at times $0,\rmd t',t'$, i.e.
\begin{align}
\star=&\int\rmd\z\int\rmd\z'U(\z)V(\z')\times\nonumber\\
&\int\rmd\beps\,\mathbb P(\beps)\varepsilon_k G(\z',t'-\rmd t'|\z+\rmd\x_0(\z,\beps))\ps(\z),
\end{align}
where the probability $\mathbb P(\beps)$ of $\sqrt{2\f D(\z)}\rmd\f
W_{\f 0}=\beps$ is given by a Gaussian distribution with zero mean and
covariance matrix $2\f D(\z)\rmd t'$. Since this distribution
is symmetric around $\f 0$, only terms with even powers of the
components of $\beps$  survive the $\mathbb P(\beps)$-integration.
Noting that for $\rmd t'\to 0$ we have $G(\z',t'-\rmd
t'|\z+\rmd\x_0(\z,\beps))\to
[1+\rmd\x_0(\z,\beps)\cdot\nabla_\z]G(\z',t'|\z)$, we see that the
only even power of the components of $\beps$ in $\varepsilon_k
G(\dots)$ gives
\begin{align}
\star=&\int\rmd\z\int\rmd\z'U(\z)\ps(\z)V(\z')\times\nonumber\\
&\int\rmd\beps\,\mathbb P(\beps)\varepsilon_k\beps\cdot\nabla_\z G(\z',t'|\z),
\end{align}
which using $\int\rmd\beps\mathbb P(\beps)\varepsilon_k\varepsilon_j=2 D_{kj}(\z)\rmd t'$ yields the result
\begin{align}
\!\star=\!\!\int\!\!\rmd\z\!\!\int\!\!\rmd\z'U(\z)\ps(\z)V(\z')\left[2\f D(\z)\nabla_\z G(\z',t'|\z)\right]_k\rmd t'.\label{lemma}
\end{align}
Rewritten terms of $P_\z(\z',t')\equiv G(\z',t'|\z)\ps(\z)$ and $\bj^g_\z\equiv-\ps(\z)\f D(\z)\nabla_\z\ps^{-1}(\z)$ we have $\bj^g_\z P_\z(\z',t')=-\ps(\z)\f D(\z)\nabla_\z G(\z',t'|\z)$, and thus
\begin{align}
\star=-2\int\rmd\z\int\rmd\z'U(\z)V(\z')\left[\bj^g_\z\right]_kP_\z(\z',t')\rmd t'.\label{lemma jg}
\end{align}
Motivated by the dual-reversal symmetry and the anticipated
applications we define the dual-reversed current operator by inverting $\bj$ and concurrently inverting $\js\to-\js$, i.e.\
\begin{align}
\revj_\x&\equiv-\bj_\x^{-\js}=-\left[\bj^g_\x-\ps^{-1}(\x)\js(\x)\right]\nonumber\\
&=\ps(\x)\f D(\x)\nabla_\x\ps^{-1}(\x)+\ps^{-1}(\x)\js(\x).
\label{current_operator_decomposed_dual}
\end{align}
Since $\revj_\z-\bj_\z=-2\bj^g$ we can rewrite Eqs.~\eqref{lemma}-\eqref{lemma jg} as
\begin{align}
\star=\int\rmd\z\int\rmd\z'U(\z)V(\z')\left(\revj_\z-\bj_\z\right)_k P_\z(\z',t')\rmd t',
\label{lemma jrev} 
\end{align}
which will turn out to be the crucial part of the following
calculations and will allow for an intuitive interpretation of the results in terms
of dual-reversed dynamics. 
 
\subsection{Application of the Lemma to initial- and  final-point currents}
In order to quantify and understand the density-current correlation
expression in Eq.~\eqref{correlation ansatz}, we now turn back to the
initial- and final-point currents, recalling the definitions in
Eq.~\eqref{IPC_FPC}. 
These observables characterize the mean initial- and final
displacements of ``pinned'' paths of duration $t_2-t_1$ conditioned on the
respective initial and final points $\z,\z'$ or $\z',\z$. The fact that both are
currents in $\z$ justifies the name ``initial- and final-point current''. Such objects turn out to play a crucial role in
the evaluation and understanding of correlations of densities and
currents, see Eq.~\eqref{correlation ansatz}. The computation of
current variances in fact involves the expectation of scalar products
of such displacements (see Eq.~\eqref{current-current ansatz}), but we
first focus on simple displacements.

Final-point currents can be computed by substituting for $\circ\rmd\x_\tau$ and integrating by parts as in Eq.~\eqref{mean_j_part2},
\begin{align}
\frac{\E{\circ\rmd\x_{t_2}}^{\x_{t_2}=\z}_{\x_{t_1}=\z'}}{\rmd t_2}=&\int\rmd\z_1\int\rmd\z_2\delta(\z_1-\z')\delta(\z_2-\z)\times\nonumber\\
&\quad P_{\z_1}(\z_2,t_2-t_1)[\f F(\z_2)+\nabla_{\z_2}^T\f D(\z_2)]\nonumber\\
=&[\f F(\z)-\f D(\z)\nabla_\z]P_{\z'}(\z,t_2-t_1)\nonumber\\
=&\bj_\z P_{\z'}(\z,t_2-t_1),
\label{FPC calculation} 
\end{align}
where the It\^o term involving $\rmd\f W_{t_2}$ vanishes whereas the Stratonovich correction term
survives. Therefore, the final-point current is obtained from the
two-point density and current operator, both appearing in the
Fokker-Planck equation (recall that $(\partial_t+\nabla_{\x}\cdot\bj_{\x})P_\y(\x,t)=0$) 
\begin{align}
\f j_{\rm fi}(\z,t_2-t_1;\z')=\bj_\z P_{\z'}(\z,t_2-t_1).\label{FPC_result}
\end{align}
For the initial-point current analogous computations yield an It\^o
increment as a correction 
\begin{align}
\f j_{\rm in}(\z',t_2-t_1;\z)&=\bj_\z P_{\z}(\z',t_2-t_1)\nonumber\\&+\langle{\sqrt{2\f D(\x_{t_1})}\rmd\f W_{t_1}}\rangle^{\x_{t_2}=\z'}_{\x_{t_1}=\z}.
\end{align}
Note that the latter It\^o increment also appears in the calculations
in Eqs.~\eqref{mean_j_part2} and \eqref{FPC calculation}, but its mean
vanishes since it involves end-point increments $\rmd\f W_{t_2}$
(note $t_2$ and not $t_1$), which are by construction uncorrelated
with the evolution up to time $t_2$. The correction term here does
\emph{not} vanish since the increment at time $t_1$ is correlated with
the probability to reach $\z'$ at time $t_2$.  Therefore this
expectation is non-trivial, but fortunately we solved this problem with the Lemma derived in
Eqs.~\eqref{expectation term}-\eqref{lemma jrev}. 

When $U$ and $V$ in Eq.~\eqref{lemma jrev} tend to a Dirac delta
function (which is mathematically not problematic since we later integrate over $\z,\z'$)
we obtain
\begin{align}
\langle{\sqrt{2\f D(\x_{t_1})}\rmd\f W_{t_1}}\rangle^{\x_{t_2}=\z'}_{\x_{t_1}=\z}=\left(\revj_\z-\bj_\z\right)P_\z(\z',t_2-t_1),
\end{align}
which gives, recalling Eq.~\eqref{current_operator_decomposed_dual}, 
\begin{align}
\f j_{\rm in}(\z',t_2-t_1;\z)=\revj_\z P_{\z}(\z',t_2-t_1).\label{IPC_result}
\end{align}
Note that $\f j_{\rm in}(\y,t;\x,0)=-\f j^{-\js}_{\rm fi}(\x,t;\y,0)$
in agreement with dual-reversal symmetry.

\begin{figure}[ht!!]
  \centering
  \includegraphics[width=.5\textwidth]{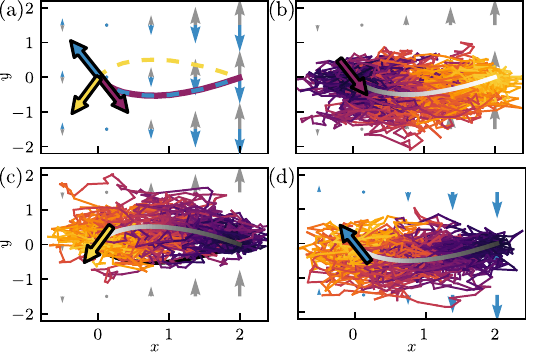}
  \caption{(a) Shear drift (gray background arrows) and inverted shear
    drift (blue background arrows) as described in the text, and
    currents and paths from (b-d) shown in purple, yellow and blue. We
    see that the purple arrow equals the inverted blue arrow, and the
    purple line overlaps with the blue dashed line, as implied by
    Eq.~\eqref{mean_paths_agree}. (b) Simulated trajectories in the
    shear flow (gray background arrows) from $\z=(0,0)^T$ to
    $\z'=(2,0)^T$ in time $t'=1$ with time always running from dark to
    bright. The initial-point current, i.e.\ the initial-point
    increment averaged over all trajectories, is depicted by the
    purple arrow and the mean paths (averaged over all trajectories)
    by the grey curve. (c) As in (b) but from $\z'=(2,0)^T$ to
    $\z=(0,0)^T$ and final-point current depicted by a yellow
    arrow. (d) As in (c) but with the inverted shear flow depicted by
    blue arrows in the background.}
 \label{fig:PRRfigIPC}
\end{figure}

To better understand these currents and their symmetry we require some
intuition about the generalized time-reversal symmetry (i.e.\ the
dual-reversal symmetry), which we gain on the basis of a simple
overdamped shear flow in Fig.~\ref{fig:PRRfigIPC}.  Consider an
isotropic diffusion with additive noise in a shear flow  $\rmd
\x_\tau=\f F_{\rm sh}(\x_\tau)\rmd\tau+\sqrt{2}\rmd \f W_\tau$ with
$\f F_{\rm sh}((x,y)^T)=(0,2x)^T$ (see gray arrows in
Fig.~\ref{fig:PRRfigIPC}a-c). \blue{For simplicity we here only
  consider shear flow in a flat potential, such that strictly
  speaking a steady-state density $\ps$ does not exist. The existence
  of $\ps$ is in fact not necessary for the discussion in this
  section, nor to connect this example to a genuine non-equilibrium
  steady state. One may equally consider the shear flow to be confined
  in a box that is large enough to allow neglecting boundary effects
  at times before $t$ and yet would yield flat $\ps$ as $t\to \infty$.} The \blue{drift of the unconfined shear flow }is purely irreversible,
i.e.\ $\f F_{\rm sh}^{\rm rev}(\x')=\f 0$. Thus, inverting the
irreversible part completely inverts the drift $\f F_{\rm
  sh}^{-\js}(\x')=-\f F_{\rm sh}(\x')$, see blue arrows in
Fig.~\ref{fig:PRRfigIPC}a,d. The initial-point current (purple arrow
in Fig.~\ref{fig:PRRfigIPC}b) is difficult to understand, since it
correlates with the constraint to reach the end point after time $t'$.
In the case of detailed balance, the time-reversal symmetry would
allow to obtain this initial-point current as the inverted final-point
current (yellow arrow in Fig.~\ref{fig:PRRfigIPC}c). However, since
detailed balance is broken by the shear flow this does not
suffice. Instead, one has to consider the final-point current for the
dynamics with the inverted irreversible drift (blue arrow in
Fig.~\ref{fig:PRRfigIPC}d). According to $\f j_{\rm in}(\y,t;\x,0)=-\f
j^{-\js}_{\rm fi}(\x,t;\y,0)$ and as can be seen in
Fig.~\ref{fig:PRRfigIPC}a, this allows to obtain the cumbersome  initial-point current (yellow) as the inverted final point current (blue).

In addition to the initial- and final-point currents, we also depict in
Fig.~\ref{fig:PRRfigIPC} the mean ``pinned'' paths. In
Fig.~\ref{fig:PRRfigIPC}a we see that the forward and
dual-reversed paths (purple and blue dashed lines) overlap. This can
also be seen from the dual-reversal symmetry in Eq.~\eqref{dual_rev}.

To prove the equality of mean paths consider $0<\tau<t'$ where $t'=t_2-t_2>0$. The (non-random) point $\boldsymbol\mu(\tau)\equiv\E{\x_{t_1+\tau}}^{\x_{t_2}=\z'}_{\x_{t_1}=\z}$ on the mean path $\z\to\z'$ is given by an integral over all possible intermediate points $\boldsymbol\mu(\tau)=\x$ weighted by $G(\z',t'-\tau|\x)G(\x,\tau|\z)/G(\z',t'|\z)$ (since $\x_\tau$ is a Markov process) which gives the Chapman-Kolmogorov-like equation
\begin{align}
G(\z',t'|\z)\boldsymbol\mu(\tau)=\int\rmd\x G(\z',t'-\tau|\x)G(\x,\tau|\z)\x.
\end{align}
The corresponding point on the mean dual-reversed path
$\boldsymbol\mu^\ddag(\tau)\equiv\langle{\x_{t_2-\tau}^{-\js}}\rangle^{\x_{t_2}=\z}_{\x_{t_1}=\z'}$
from $\z'$ to $\z$ with reversed steady-state current $\js\to-\js$ is
given by (using three times the dual-reversal in Eq.~\eqref{dual_rev})
\begin{align}
&G^{-\js}(\z,t'|\z')\boldsymbol\mu^\ddag(t'-\tau)\nonumber\\
&=\int\rmd\x G^{-\js}(\z,\tau|\x)G^{-\js}(\x,t'-\tau|\z')\x\nonumber\\
&=\int\rmd\x G(\x,\tau|\z)\frac{\ps(\z)}{\ps(\x)}G(\z',t'-\tau|\x)\frac{\ps(\x)}{\ps(\z')}\x\nonumber\\
&=\frac{\ps(\z)}{\ps(\z')}G(\z',t'|\z)\boldsymbol\mu(\tau)\nonumber\\
&=G^{-\js}(\z,t'|\z')\boldsymbol\mu(\tau),\label{mean_paths_agree} 
\end{align}
which implies $\boldsymbol\mu(\tau)=\boldsymbol\mu^\ddag(t'-\tau)$ for
all $t_1<\tau<t_2$, so the mean paths indeed agree (but  run in
opposite directions), which completes the proof that the blue and purple paths in Fig.~\ref{fig:PRRfigIPC}a overlap.
\subsection{Current-density correlation}
With the definitions \eqref{IPC_FPC} and $t'=t_2-t_1>0$ we have
(recall the simplification of $\integrals_{\x\y}^{t,U}$ in Eq.~\eqref{int_op_simpl})
\begin{align}
\f C^{\x\y}_{\f J \rho}(t)=\integrals_{\x\y}^{t,U}[\f j_{\rm fi}(\z,t';\z')+\f j_{\rm in}(\z',t';\z)-2\js(\z)\ps(\z')].
\end{align}
As we have shown in Eqs.~\eqref{FPC_result} and \eqref{IPC_result} the
initial- and final-point currents can be expressed in terms of the
current operators yielding 
\begin{align}
\f C^{\x\y}_{\f J \rho}(t)=\integrals_{\x\y}^{t,U}[\bj_\z
  P_{\z'}(\z,t')+\revj_\z P_\z(\z',t')-2\js(\z)\ps(\z')],
\label{current-density-result} 
\end{align}
which allows to explicitly calculate $\f C^{\x\y}_{\f J \rho}(t)$ if $P_{\z'}(\z,t')$ is known. An analogous result for the scalar current variance was very recently obtained in
\cite{Dechant2021PRR} but did not establish a connection to current
operators and dual-reversal symmetry and did not consider coarse
graining nor multi-dimensional continuous-space examples. The current-density correlation $\f C^{\x\y}_{\f J \rho}(t)$ can be interpreted analogous to $\f C^{\x\y}_{\rho\rho}(t)$ as follows.

All possible paths between points $\z,\z'$ in time $0<t'\le t$
contribute, weighted by their corresponding probability, to this
correlation.  The difference with respect to density correlations $\f
C^{\x\y}_{\rho\rho}(t)$ is that now currents at position $\z$ are
correlated with probabilities to be at the point $\z'$. For
paths $\z'\to\z$ the displacement is obtained from the familiar
current operator $\f j_{\rm fi}=\bj_\z P_{\z'}(\z,t')$. Paths from
$\z\to\z'$ are mathematically more involved (and somewhat harder to
understand), but can be understood intuitively with the dual-reversal
symmetry (see also Fig.~\ref{fig:PRRfigIPC}).  More precisely, they
can be understood and calculated in terms of the dual-reversed current
operator $\revj_\z\equiv-\bj_\z^{-\js}$. 

A direct observation that follows from the result in
Eq.~\eqref{current-density-result} is that at equilibrium (i.e.\ under detailed
balance), we have $\js=\f 0$,\ \,$\revj_\z=-\bj_\z$ and
$P_\z(\z',t')=P_{\z'}(\z,t')$ and thus $\f C^{\x\y}_{\f J \rho}(t)=\f 0$
for all window functions and all points $\x,\y$.
The correlation $\f C^{\x\y}_{\f J \rho}(t)$ can also be utilized to
improve the thermodynamic uncertainty relation (TUR), as recently shown in
\cite{Dechant2021PRX}. The result in
Eq.~\eqref{current-density-result} thus allows to inspect and
understand more deeply this improved TUR.

\begin{figure}[ht!!]\begin{center}\includegraphics[width=.45\textwidth]{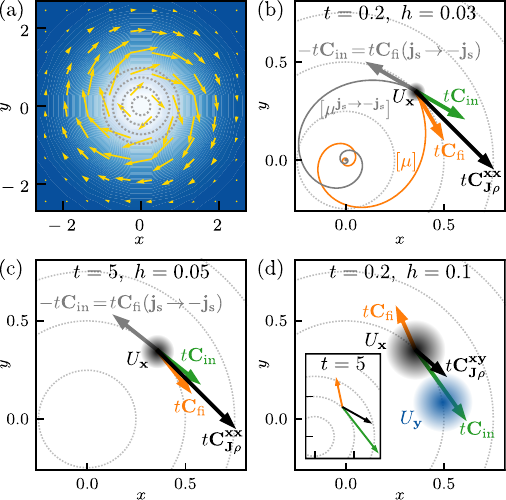}\caption{(a)
      Illustration of the steady-state density (color gradient) and
      current (arrows) of the two-dimensional rotational flow
      Eq.~\eqref{OUP} with $\Omega=3$. Gray dotted lines in (a-d) are
      circles with radii $0.25,\,0.5,\,0.75,\,1$. (b-c) Single-point
      $\x=\y$ and (d) two-point time-accumulated correlation $t\f
      C^{\x\y}_{\f J \rho}$ at $t=0.2$ and $t=5$ (black arrow), with
      final-point $\f C_{\rm fi}\equiv \integrals^t_{\x\y}[\bj_\z
        P_{\z'}(\z,t')-\js(\z)\ps(\z')]$ (orange) and initial-point
      $\f C_{\rm in}$ (green) contribution, s.~t. $\f C^{\x\y}_{\f J \rho}=\f C_{\rm in}+\f C_{\rm fi}$. $\f C_{\rm fi}(\js \to -\js)$ (gray) is the current-reversed final-point contribution which agrees with the inverted initial-point contribution $-\f C_{\rm in}$. Full lines in \blue{(b)} are the mean trajectory $[\mu]\equiv\langle\x_{\tau\ge0}\rangle_{\x_0=\x}$ (orange) and its current-reverse $[\mu^{\js\to-\js}]$ (gray). $U_{\x,\y}$ (shaded circles) is a Gaussian at $\x,\y$ with width $h$, see Eq.~\eqref{Gauss window}.\label{fig:correlation}}\end{center}\end{figure}

An explicit example of the correlation result Eq.~\eqref{current-density-result} for $\f
C^{\x\y}_{\f J \rho}(t)$ is shown in Fig.~\ref{fig:correlation}. In line with the previous arguing $\f
C^{\x\y}_{\f J \rho}(t)$ can be understood 
as a vector with initial- and final-point contributions, $C^{\x\y}_{\f J \rho}=\f C_{\rm in}+\f C_{\rm fi}$, where $\f C_{\rm in}\equiv \integrals^t_{\x\y}[\revj_\z P_{\z}(\z',t')-\js(\z)\ps(\z')]$. 
In the Supplemental Material of \cite{AccompanyingLetter} we have shown that for $\x=\y$ in the limit $h\to 0$
of small windows the results for the correlation simplify $\f C_{\rm
  in}(t)\simeq[2\js(\x)/\ps(\x)-\f F(\x)]\text{var}^\x_\rho(t)/4$ and
$\f C_{\rm fi}(t) \simeq {\f F(\x)}\text{var}^\x_\rho(t)/4$, implying $\mathbf{C}^{\x\x}_{\mathbf{J}\rho}(t)\simeq
\js(\x)\text{var}^\x_\rho(t)/2\ps(\x)$. Since $\f F=\f F^{\rm rev}+\js/\ps$ and thus $2\js(\x)/\ps(\x)-\f F(\x)=-\f F^{-\js}(\x)$, the above implies that for $\x=\y$ and small windows $h$ we have $-\f C_{\rm in}=\f C^{-\js}_{\rm fi}$ and $\f C_{\rm fi}$ points along $\f F(\x)$ that is tangent to the mean trajectory $[\mu]$ at $\x$, while $\mathbf{C}^{\x\x}_{\mathbf{J}\rho}(t)$ points in $\js(\x)$-direction, see Fig.~\ref{fig:correlation}b. For longer times $t$ and/or larger $h$, the direction of $\f C_{\rm fi}$ changes but $-\f C_{\rm in}=\f C^{-\js}_{\rm fi}$ still holds (see Fig.~\ref{fig:correlation}c) since the symmetry $\f j_{\rm in}(\y,t;\x,0)=-\f j^{-\js}_{\rm fi}(\x,t;\y,0)$ can be applied in the integrands.
Conversely, the two-point correlation
$C^{\x\y}_{\mathbf{J}\rho}$ \emph{need not} to point along $\js(\x)$
(Fig.~\ref{fig:correlation}d). In fact, its direction changes over
time (see inset of
Fig.~\ref{fig:correlation}d). Notably, results for $\x\ne\y$ akin to
Fig.~\ref{fig:correlation}d may provide deeper insight into barrier 
  crossing problems on the level of individual trajectories in the absence of detailed balance.

\subsection{Current (co)variance}\label{subsec:current-current}
Recall that the current (co)variance Eq.~\eqref{current-current ansatz} involves
 scalar products of initial- and final-point increments
 ${\E{\circ\rmd\x_{t_1}\cdot\circ\rmd\x_{t_2}}}^{\x_{t_2}=\z'}_{\x_{t_1}=\z}$,
 which \emph{cannot} be easily interpreted as scalar products of
 currents.  They are \emph{not} the scalar products of initial- and
 final-point currents, since
 ${\E{\circ\rmd\x_{t_1}\cdot\circ\rmd\x_{t_2}}}^{\x_{t_2}=\z'}_{\x_{t_1}=\z}\ne\E{\circ\rmd\x_{t_1}}^{\x_{t_2}=\z'}_{\x_{t_1}=\z}\cdot\E{\circ\rmd\x_{t_2}}^{\x_{t_2}=\z'}_{\x_{t_1}=\z}$. Rather
 they correspond to the scalar product of the initial- and
 final-point increment \emph{along the same trajectory} and only then
 they become averaged over all trajectories from $\z$ to $\z'$ (see
 also Fig.~2 in \cite{AccompanyingLetter}). For $t_1<t_2$ these are computed equivalently
 to Eqs.~\eqref{FPC calculation}-\eqref{IPC_result} based on the Lemma \eqref{lemma jrev} as 
\begin{align}
\E{\circ\rmd\x_{t_1}\cdot\circ\rmd\x_{t_2}}^{\x_{t_2}=\z'}_{\x_{t_1}=\z}=\revj_\z\cdot\bj_{\z'}P_\z(\z',t').\label{scalar product increments result}
\end{align}
However, according to the convention $\int_{t_1}^t\rmd
t_2\delta(t_2-t_1)=1/2$ in Eq.~\eqref{int_op}, we also need to
consider the case $t_1=t_2$, i.e.\ $t'=0$, which did not contribute
for $C_{\rho\rho}$ and $C_{\f J\rho}$. In the case $t_1=t_2$ (recall
the definition in Eq.~\eqref{notation_end-points})
\begin{align}
\langle \circ\rmd\x_{t_1}&\cdot\circ\rmd\x_{t_2}\rangle^{\x_{t_2}=\z'}_{\x_{t_1}=\z}\nonumber\\ 
\equiv&\Es{\delta(\x_{t_1}-\z)\circ\rmd\x_{t_1}\cdot\delta(\x_{t_2}-\z')\circ\rmd\x_{t_2}}\nonumber\\
\overset{t_1=t_2}=&\Big\langle \delta(\x_{t_1}-\z)\sqrt{2\f D(\x_{t_1})}\rmd\f W_{t_1}\cdot \nonumber\\
&\,\,\,\delta(\x_{t_2}-\z')\sqrt{2\f D(\x_{t_2})}\rmd\f W_{t_2}\Big\rangle,
\end{align}
where we used that for $t_1=t_2$ the only term surviving is $\rmd
\f W_{t_1}^{\,2}$ (and not $\rmd\f W_{t_1}\rmd t_1$ and $\rmd t_1^2$, which is
why such terms only enter in current-current expressions but not in current-density or
density-density correlations),  as well as (by It\^o's isometry)
$\int_{t_1}^t\rmd t_2\rmd\f W^j_{t_1}\frac{\rmd\f W^l_{t_2}}{\rmd t_2}=\delta_{jl}\rmd t_1$. Using
$P_\z(\z',t'=0)=\delta(\z-\z')\ps(\z)$ we find for $t_1=t_2$
\begin{align}
&\E{\circ\rmd\x_{t_1}\cdot\circ\rmd\x_{t_2}}^{\x_{t_2}=\z'}_{\x_{t_1}=\z}\nonumber\\
&=P_\z(\z',0)\sum_{i,j,l=1}^d[\sqrt{2\f D(\z)}]_{ij}[\sqrt{2\f D(\z')}]_{il}\delta(t_1-t_2)\delta_{jl}\rmd t_1\nonumber\\
&=\ps(\z)\delta(\z-\z')\sum_{i=1}^d[2\f D(\z)]_{ii}\delta(t_1-t_2)\rmd t_1\nonumber\\
&=2{\rm Tr}[\f D(\z)]\ps(\z)\delta(\z-\z')\delta(t_1-t_2)\rmd t_1.\label{same time part}
\end{align}
Plugging this into Eq.~\eqref{current-current ansatz},
we obtain, using  Eq. \eqref{scalar product increments result} and
accounting for the $t'=0$ contribution, the result for current
covariances in the form of
\begin{align}
&C^{\x\y}_{\f J\cdot \f J}(t)=\frac2t\int d\z {\rm Tr}[\f D(\z)]U^h_\x(\z)U^h_\y(\z)\ps(\z)\label{current covariance result}\\
&+\integrals_{\x\y}^{t,U}[\revj_{\z'}\cdot\bj_\z P_{\z'}(\z,t')+\revj_\z\cdot\bj_{\z'}P_\z(\z',t')-2\js(\z)\cdot\js(\z')].\nonumber
\end{align}
The second line is interpreted analogously to the current-density
correlation in Eq.~\eqref{current-density-result} with the only
difference that the scalar product of current operators reflects
scalar products of increments along individual trajectories.  The
first term, however, does not appear in $C^{\x\y}_{\f J\rho}$ and
$C^{\x\y}_{\rho\rho}$. As can be seen from the derivation in Eq.~\eqref{same time part} this term originates from
the purely diffusive (i.e. Brownian) term involving
$\rmd\x_\tau\cdot\rmd\x_\tau=2{\rm Tr}\f D(\z)\rmd\tau$ and only
appears for $t_1=t_2$, i.e.\ $t'=0$.  Thus,  this term \emph{cannot}
be interpreted in terms of trajectories from $\z$ to $\z'$ or vice
versa, but instead reflects that due to the nature of Brownian motion
the square of instantaneous fluctuations $(\rmd\x_\tau)^2$ does not
vanish but contributes on the order $\rmd\tau$.  Note that since here
$\z=\z'$ this term only contributes if $U^h_\x(\z)$ and $U^h_\y(\z')$
have non-zero overlap.

For $\x=\y$ the covariance becomes the current variance ${\rm var}_{\f J}^\x(t)\equiv C^{\x\y}_{\f J\cdot \f J}(t)$ which plays a vital
role in stochastic thermodynamics.  As an application of the result
in Eq.~\eqref{current covariance result} we use the 
TUR-bound under concurrent variation of the coarse-graining scale $h$
to optimize the inference of a system's dissipation via current
fluctuations. \blue{Before we turn to this inference problem, we take a closer look at the limit of no coarse graining, i.e.\ $h\to 0$.}

\section{The limit of no coarse graining}\label{sec:h to 0}
\blue{In this section we consider the variance ${\rm
    var}_{\rho}^\x(t)\equiv C^{\x\x}_{\rho\rho}(t),{\rm var}_{\f
    J}^\x(t)\equiv C^{\x\x}_{\f J\cdot \f J}(t)$ and correlations $\f
  C^{\x\x}_{\f J \rho}(t)$ in Eqs.~\eqref{density-density
    result},\eqref{current-density-result},\eqref{current covariance
    result} with $\x=\y$ in the limit of no coarse graining,
  i.e.\ when $h\to 0$. In particular, we consider normalized window
  functions $\int\rmd\z U^h_\x(\z)=1$ such that in the limit of no
  coarse graining $U^{h\to 0}_\x(\z)=\delta(\x-\z)$ (see
  e.g.\ \eqref{Gauss window}). Thus, the density and current
  observables in Eq.~\eqref{def_current} for $h=0$ correspond to the empirical density and current defined with a delta function
\begin{align} 
\overline{\rho_\x}(t)&\equiv \frac{1}{t}\int_0^t\delta(\x-\x_\tau)\rmd\tau\nonumber\\
\overline{\f J_\x}(t)&\equiv \frac{1}{t}\int_{\tau=0}^{\tau=t}\delta(\x-\x_\tau)\circ\rmd\x_\tau,
\label{def_current_delta}
\end{align}
which is the definition typically adopted in the literature \cite{Maes2008PA,Touchette2009PR,Kusuoka2009PTRF,Chetrite2013PRL,Chetrite2014AHP,Barato2015JSP,Hoppenau2016NJP,Touchette2018PA,Mallmin2021JPAMT,Monthus2021JSMTE}.
We show in Appendix~\ref{sec:app} that in spatial dimensions $d\ge 2$ the variance and correlation functions diverge ${\rm var}_{\rho}^\x(t),{\rm var}_{\f J}^\x(t),\f C^{\x\x}_{\f J \rho}(t)\to\infty$ as $h\to 0$. Note that the mean values Eqs.~\eqref{mean_rho} and \eqref{mean_j_part2} of the observables Eq.~\eqref{def_current_delta} do not diverge but instead for $U^{h\to 0}_\x(\z)=\delta(\x-\z)$ directly simplify to $\Es{\overline{\rho_\x}(t)}=\ps(\x)$ and $\Es{\overline{\f J_\x}(t)}=\js(\x)$ (see also \cite{Maes2008PA}).

Before we go into the specific results for the limit $h\to 0$, let us
first discuss why divergent fluctuations of the functionals in
Eq.~\eqref{def_current_delta}, although overlooked so far,  are in
fact \emph{not} surprising. The simplest argument is that second
moments as e.g.\ $\Es{\overline{\rho_\x}(t)^2}$ involve terms
$\Es{\delta(\x-\x_\tau)\delta(\x-\x_{\tau'})}$, which diverge for
$\tau=\tau'$ since a squared delta function appears. In contrast, the
mean value  $\Es{\overline{\rho_\x}(t)}$ contains
$\Es{\delta(\x-\x_\tau)}=\ps(\x)$ which is finite. Loosely speaking,
the mean value involving $\Es{\delta(\x-\x_\tau)}$ is given by the
probability to be at point $\x$, which is zero, multiplied by the
height of the delta function at $\x$, which is infinite. Since the
mean value is finite for $h\to 0$ this can be seen to yield
``$0\times\infty=\ps(\x)$'', while the second  moment contains a squared
delta peak, such that the second moment loosely speaking diverges due
to ``$0\times\infty^2=\ps(\x)\times\infty=\infty$''.  This argument
illustrates that divergent fluctuations are not surprising but this
argument is oversimplified since it does not take into account the
time integration.  In particular, to explain why the divergence only
occurs in spatial dimensions $d\ge 2$, we have to note that due to the
time integration  the one-dimensional case is qualitatively different.
Given some point $\z$ in $d$-dimensional space, the trajectory will
hit $z\equiv\z$ with a finite probability in $d=1$ (i.e.\ with
non-zero probability there is some $\tau\in[0,t]$ such that
$x_\tau=z$; e.g.\ if $x_0<x_t$ all points in $[x_0,x_t]$ are
hit). This is qualitatively different for $d\ge 2$,  since overdamped
motion in $d\ge 2$ does not hit points, i.e.\ the probability to hit a
given point $\z$ is zero,
$\mathbbm{P}(\exists\tau\in(0,\infty)\colon\x_{\tau}=\z)=0$
\cite{Durrett_Stoch}. This property is not specific to overdamped
motion, but is rather due to the fact that the set of points
$(\x_\tau)_{0\le\tau\le t}$ has Lebesgue measure zero for $d\ge 2$.  

To further explain the divergence and its dependence on the
dimensionality in a somewhat less oversimplified way (for the detailed
derivation see Appendix~\ref{sec:app}), we take a second look at the
term
$\Es{\delta(\x-\x_\tau)\delta(\x-\x_{\tau'})}=G(\x,\abs{\tau-\tau'}|\x)\ps(\x)$
occurring in $\Es{\overline{\rho_\x}(t)^2}$. Here, $G(\x,t'|\x)$
trivially diverges if $t'=0$. However, the relevant question is whether
the return integral $\int_0^t G(\x,t'|\x)\rmd t'$ diverges. Any
divergence in the integral would come from $t'\to 0$  where
$G(\x,t'|\x)$ diverges, i.e.\ from the limit of small time differences
$\abs{\tau-\tau'}$. For $t'\to 0$ the overdamped propagator
$G(\x,t'|\x)$ becomes Gaussian with variance $\propto Dt'$
\cite{Risken1996} (so for very small $t'$ we have $G(\x,t'|\x)\propto
     {t'}^{-d/2}$ in $d$-dimensional space), and thus the return
     integral $\int_0^t G(\x,t'|\x)\rmd t'$ diverges if and only if
     $\int_0^t {t'}^{-d/2}\rmd t'$ diverges. Therefore the variance
     ${\rm var}_{\rho}^\x(t)$ diverges in spatial dimensions $d\ge 2$.

Apart from the two arguments above providing mathematical intuition
about the divergence, there is also a physical intuition that suggests
divergent fluctuations. Recall that for finite $h>0$, the observables
$\overline{\rho^U_\x}$ and $\overline{\f J^U_\x}$ in
Eq.~\eqref{def_current} by definition measure the time and
displacement that the trajectory $(x_\tau)_{0\le\tau\le t}$
accumulates in the region $U^h_\x$ of scale $h$ around $\x$. Now as
$h\to 0$, only visitations of precisely the point $\x$ contribute.
Two very similar (but not equal) trajectories may now give very
different values for $\overline{\rho^U_\x}$ and $\overline{\f
  J^U_\x}$, depending whether the point $\x$ is hit or even slightly missed
(e.g.\ by a distance $h$).  Therefore, fluctuations among different
trajectories of these functionals diverge as $h\to 0$.
This reasoning
is not restricted to overdamped stochastic motion, and indeed seems to
hold for more general dynamics, see outlook in Sec.~\ref{sec:outlook}.

This simple illustration also explains why fluctuations do not diverge
in one-dimensional space. There, points are hit, meaning that e.g.\ a
trajectory starting at $0$ and ending at $1$ always hits all points in
between at some intermediate time, which is why the density and
current observables have qualitatively lower fluctuations compared to
higher dimensions. The reason that the divergence for $d\ge 2$ was
overlooked so far is probably due to the fact that most explicit
examples were analyzed in one-dimensional space only.

Explicitly, in the limit $h\to 0$ the expressions Eqs.~\eqref{density-density result}, \eqref{current-density-result},\eqref{current covariance result} with $\x=\y$ for any time $t$ take the form 
\begin{align}
{\rm var}^\x_\rho(t)\!\!&\overset{h\to 0}{\simeq}\frac{K}{\tilde{D}_\x t}\ps(\x)
\begin{cases} \frac{h^{2 - d}}{d - 2} & \text{for}\: d > 2 \\ -\ln{h} & \text{for}\: d = 2 \end{cases}
\nonumber\\
\f C^{\x\x}_{\f J\rho}(t)\!\!&\overset{h\to 0}{\simeq}
\js(\x){\rm var}^\x_\rho(t)/2\ps(\x) \label{bounds h to 0}\\
\text{var}^{\x}_{\f J}(t)\!\!&\overset{h\to 0}{=}\!\!K'
\frac{2\tilde{D}'_\x}{t}\ps(\x)(d-1)h^{-d}\!+\mathcal{O}(t^{-1})\mathcal{O}(h^{1-d}),\nonumber
\end{align}
where  $\simeq$ denotes asymptotic equality,
$\tilde{D}_\x,\tilde{D}'_\x$ are constants bounded by the smallest and
largest eigenvalues of $\f D(\x)$, and $K,K'$ are constants depending
on the specific normalized window $U^h_\z$ (see
Appendix~\ref{sec:app}). Note that the dominant term in
$\text{var}^{\x}_{\f J}(t)$ vanishes for $d=1$ such that all three
expressions only diverge for $d\ge 2$. Some details on the case $d=1$
are shown in Appendix~\ref{1d h=0}. 

Thus, the empirical density and current as defined in
Eq.~\eqref{def_current_delta} have divergent fluctuations. Note that
an infinite variance contradicts Gaussian statistics on all time
scales. This divergence, moreover, leads us to question whether
Eq.~\eqref{def_current_delta} is even well-defined, i.e.\ whether
these observables are mathematically well-defined random variables,
and whether the result in the limit $h\to 0$ is unaffected by the
specific choice of the $U^h_\x$ as long as $U^{h\to
  0}_\x(\z)=\delta(\x-\z)$.} 

\section{Application to inference of dissipation}\label{APPLI}
We now apply the results for the current variance ${\rm var}_{\f J}^\x(t)\equiv C^{\x\x}_{\f J\cdot \f J}(t)$ in Eq.~\eqref{current
  covariance result} for $\x=\y$. For an individual component,
e.g.\ $J_y\equiv[\overline{\f J^U_\x}]_y$, of the vector $\overline{\f J^U_\x}$ the equivalent result reads 
\begin{align}
&{\rm var}^{\x}_{J_y}(t)\!=\!\frac2t\!\int\!\!d\z [\f D(\z)]_{yy}U^h_\x(\z)U^h_\x(\z)\ps(\z)+\integrals_{\x\y}^{t,U}[(\revj_{\z'})_y
\times\nonumber\\&
(\bj_\z)_y P_{\z'}(\z,t')+(\revj_\z)_y (\bj_{\z'})_y P_\z(\z',t')-2[\js(\z)]_y[\js(\z')]_y].\label{current component variance result}
\end{align}
With the dissipation rate $\dot\Sigma$ in Eq.~\eqref{EPR}, current
observables such as $J_y\equiv[\overline{\f J^U_\x}]_y$ satisfy the
TUR \cite{Barato2015PRL,Dechant2018JSMTE} (in the form relevant below
first proven in \cite{Dechant2018JSMTE})
\begin{align}
\frac{{\rm var}^{\x}_{J_y}(t)}{\Es{J_y}^2}\ge \frac2{t\dot\Sigma}.\label{TUR} 
\end{align}
This bound is of particular interest since it allows to infer a lower
bound on a system's  dissipation from measurements of the local mean current and
current fluctuations \cite{Li2019NC,Supriya,Otsubo2020PRE,Vu2020PRL,Horowitz2019NP}. 
Note that Eq.~\eqref{TUR}
implicitly assumes ``perfect'' statistics, i.e.\ $\Es{J_y}$ and
${\rm var}^{\x}_{J_y}(t)$ are the exact  mean and variance for the
process under consideration (not limited by sampling
constraints on a finite number of realizations).

We now investigate the influence of the coarse graining on the
sharpness of the bound \eqref{TUR}. One might
naively expect that coarse graining annihilates information. However,
as shown in \cite{AccompanyingLetter}  the current fluctuations
diverge in spatial dimensions $d\ge 2$ in the limit $h\to 0$ (of no
coarse graining),  whereas the mean converges to a constant (note that
$\dot\Sigma$ does not at all depend on $U^h_\x$). The exact
asymptotics for $h\to 0$ in
\cite{AccompanyingLetter} 
demonstrate that the bound \eqref{TUR} becomes entirely independent of the process
(i.e.\ it only depends on $\ps$ but contains no information about
the non-equilibrium part of the dynamics). Therefore, the left hand side of
the inequality \eqref{TUR} tends to $\infty$ as $h\to 0$, rendering the TUR without
spatial coarse graining unable to infer dissipation beyond the
statement $\dot\Sigma\ge 0$ for $h=0$.

However, the naive intuition is correct in the limit of ``ignorant'' coarse graining $h\to\infty$, where $U^h_\x$ becomes asymptotically constant in a
sufficiently large hypervolume centered at $\x$ (i.e.\ in a
hypervolume ${\rm A}$ where $\int_{\rm A}\ps(\x)\rmd\x\approx 1$). The
integration over a constant $U^h_\x=c$ yields $\Es{\overline{\f J^U_\x}(t)}=c\int\rmd\z \js(\z)=\f 0$  for the
mean Eq.~\eqref{mean_j_part2}. The vanishing $\Es{\overline{\f J^U_\x}(t)}$ may be seen in two
  ways.
  First, since $\nabla_\z\cdot\js(\z)=0$, curl
  $\js(\z)=\nabla_\z\times\f f(\z)$ and by Stokes theorem $\int_{\rm
    A}\rmd^2z\nabla_\z\times\f f(\z)=\int_{\partial\rm A}\f
  f\cdot\rmd\f l$ which vanishes since at the boundary $\partial{\rm A}$
  at $\infty$ we have $\ps\to 0$, thus $\js\to 0$ and therefore the vector
  potential $\f f\to\f 0$. Second,  for $U^h_\x=c$ we have
  $\overline{\f J^U_\x}(t)=\frac{c}{t}(\f x_t-\f x_0)$ (and we assume
  $\f x_0$ to be sampled from $\ps(\x)$). Then $\f x_0$ and $\f x_t$ are both distributed according to $\ps$, thus $\Es{\f x_t}=\Es{\f x_0}$ and $t\Es{\overline{\f J^U_\x}(t)}/c=\Es{\f x_t}-\Es{\f x_0}=0$.
Conversely, the variance remains strictly
  positive. Therefore, also for $h\to\infty$ the left hand side of the inequality
 \eqref{TUR} diverges, rendering the TUR with an ``ignorant''
coarse graining incapable of inferring dissipation (again only gives $\dot\Sigma\ge 0$ as for $h=0$). 

These two arguments, i.e.\ the necessity of coarse graining
\cite{AccompanyingLetter} and the failure of an ``ignorant'' coarse
graining, imply that an intermediate coarse graining exists that is optimal
for inferring dissipation via the TUR \eqref{TUR}. 

\begin{figure*}[ht!!]\begin{center}
\includegraphics[width=0.9\textwidth]{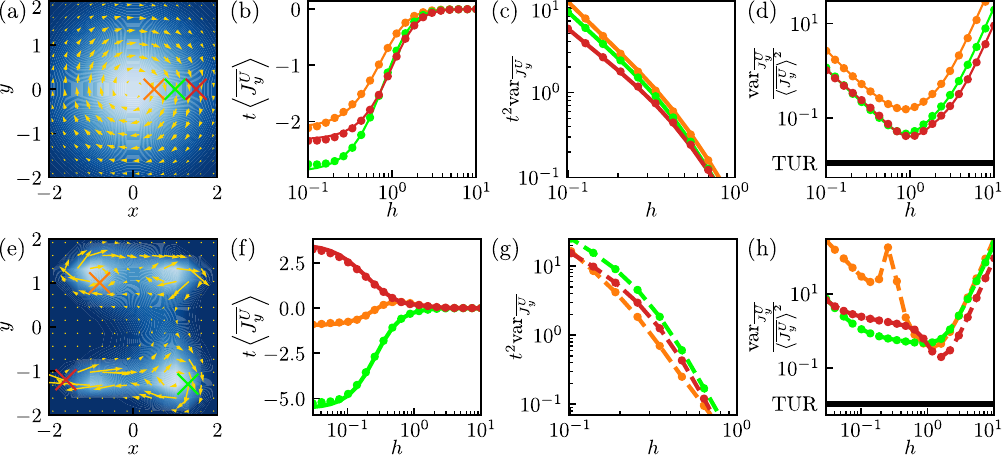}
\caption{
(a) Steady-state density (blue color gradient) and current (yellow
  arrows) for the rotational flow Eq.~\eqref{OUP} with $\Omega=3$. Points around which the currents are evaluated in
  (b-d) are denoted by colored crosses.  (b) Simulated values (circles)
  of the mean $y$-component of the time-integrated current from
  $2,000$ trajectories of length $t=10$ with time-step $dt=0.001$
  (using the stochastic Euler algorithm) starting from steady-state initial
  conditions using a Gaussian window function Eq.~\eqref{Gauss window} with
  different coarse-graining scales $h$.  Analytical results
  Eq.~\eqref{mean_j_part2} are shown with lines. (c) As in (b) but for
  variances. Simulations (circles) are shown alongside
  analytical results (lines;
  the results are analytic up  to one time-integration, see Eq.~\eqref{current component variance result}) for the variance of currents. (d) The relative error (ratio of variance and mean-squared)
  as a function of $h$ features a minimum at an intermediate $h$. At
  this minimum, the current fluctuations give the best lower bound on
  the dissipation via the TUR  \eqref{TUR} at the value
  $2/(t\dot\Sigma)=2/(10\times 18)=0.011$   (black line).
(e) As in (a) but for the more complicated process in Eq.~\eqref{SDE_4well}
  with $D=1$ and we choose $\Omega=0.957$ to have the same dissipation
  as in (d); here the dissipation is obtained by means of a
  numerical integration. (f) As in (b) but for the process in (e) the
  ``analytical'' mean \eqref{mean_j_part2} had to be evaluated by
  means of a
  numerical integration. (g) As in (c) but simulated values are shown
  by circles and dashed line (but without a comparison to results of numerical
  integration since these require the knowledge of the
  propagator). (h) As in (d) but for the process in (e). The relative error 
  may display several local minima. Some intermediate $h$ still allows
  for an optimal inference 
  of the dissipation via the TUR (black line). Note that the relative
  error diverges (orange line) where the mean crosses zero (orange
  line in (b)). 
\label{fig:TUR}}\end{center}\end{figure*}

We first demonstrate this finding using a two-dimensional rotational flow \eqref{OUP} with Gaussian coarse graining window Eq.~\eqref{Gauss window}. We evaluate the left hand side of Eq.~\eqref{TUR} for varying $h$ and
$\x$ and compare it to the constant right hand side of
Eq.~\eqref{TUR}. Particularly for $\f D(\z)=D\f 1$, we have $\ps(\z)=r/(2\pi
D)\exp(-r\z^2/(2D))$ and $\js(\z)=\Omega\ps(\z)(z_2,\,-z_1)^T$ and the dissipation rate 
Eq.~\eqref{EPR} is given by
\begin{align}
\dot\Sigma&=\int\rmd\z\frac{\js^T(\z)}{\ps(\z)}\f D^{-1}(\z)\frac{\js(\z)}{\ps(\z)}\ps(\z)=\frac{\Omega^2}{D}\int\rmd\z\,\z^2\ps(\z)\nonumber\\
&=\frac{\Omega^2}{D}\Es{\x_0^2}=\frac{\Omega^2}{D}\Es{x_1^2+x_2^2}=\frac{\Omega^2}{D}2\frac{D}{r}=\frac{2\Omega^2}{r}.
\end{align}
Thus the TUR in Eq.~\eqref{TUR} for the rotational flow becomes
\begin{align}
\frac{{\rm var}^{\x}_{J_y}(t)}{\Es{J_y}^2}\ge
\frac{r}{t\Omega^2}.
\end{align}
The results shown in Fig.~\ref{fig:TUR}a-d demonstrate, as argued
above, that relative fluctuations diverge as $h\to 0,\infty$.  For
this example, the relative error as a function of $h$ has a unique
minimum (slightly depending on $\x$, and possibly on other parameters
such as $t$). This means that (restricted to $U^h_\x$ being a Gaussian
around $\x$) there is a coarse graining scale  $h$ that is optimal for
inferring a lower bound on the dissipation, that may also provide some
intuition about the formal optimization carried out in
\cite{Vu2020PRL}. This result demonstrates that coarse graining trajectory
data \emph{a posteriori} can improve the inference of thermodynamical
information, which is a strong motivation for considering coarse
graining. 

In particular, note that this method is readily applicable, i.e.\ one
does not need to know the underlying process (as long as the dynamics
is overdamped). As was done in Fig~\ref{fig:TUR}e-h one simply
integrates the trajectories to obtain the coarse grained current as
defined in Eq.~\eqref{def_current}. Then, the mean and variance are
readily obtained from the fluctuations along an ensemble of individual
trajectories, and for each value of $\x$ and $h$ one determines a
lower bound on the dissipation via Eq.~\eqref{TUR}. Finally, one takes
the best of those bounds. We here only consider Gaussian $U^h_\x$ for
the coarse graining, but due to the flexibility of the theory one could
even choose window functions that do not have to relate to the notion of coarse graining.
Notably, a Gaussian window function is in this
case better than e.g.\ a rectangular indicator function (which one
usually uses for binning data
) due to an
improved smoothing effect. Moreover, one further expects a reduced error
due to discrete-time effects.

Note that compared to many of the similar existing methods \cite{Dechant2021PRX,Li2019NC,Gingrich2017JPAMT}, we \emph{neither}
advise to rasterize the continuous dynamics to parameterize
(i.e.\ ``count'') currents \emph{nor} to
approximate the dynamics by a Markov-jump process\blue{. Our method is therefore not only correct (note that a Markov-jump assumption is only accurate in the presence of a time-scale separation ensuring a local
equilibration, e.g. as a results of high barriers separating energy
minima) but also} has the great advantage of not having to parameterize rates at
all. Instead one simply integrates trajectories according to Eq.~\eqref{def_current}.

A generalization to windows that are not centered at individual
points as well as the use of correlations
in Eq.~\eqref{current-density-result} entering the recent so-called CTUR
inequality \cite{Dechant2021PRX} will be considered in forthcoming publications.

To underscore the applicability of the above inference strategy, we
apply it to a more complicated system, for which a Markov jump process
description would be difficult due to the presence of low and flat barriers and
extended states. The results are shown Fig.~\ref{fig:TUR}e-h. The example is constructed by considering the two-dimensional potential
\begin{align}
  \phi(x,y)=&0.75 (x^2 - 1)^2 + \nonumber\\
  &(y^2 - 1.5)^2 ((x + 0.5 y - 0.5 )^2 + 0.5)+c\label{phi_4well} 
\end{align}
where $c$ is a constant such that $\ps(\z)=\exp[-\phi(\z)]$ is
normalized. We consider isotropic additive noise $\f D(\z)=D\f 1$ and
construct the It\^o/Langevin equation for the process as
\begin{align}
\rmd\x_\tau=-D\{\nabla\phi\}(\x_\tau)\rmd\tau+\f F^{\rm irrev}(\x_\tau)+\sqrt{2D}\rmd\f W_\tau,\label{SDE_4well} 
\end{align}
where
\begin{align}
\f F^{\rm irrev}(\z)=\frac{\js(\z)}{\ps(\z)}\equiv-D\Omega\begin{bmatrix}0&-1\\1&0\end{bmatrix}\cdot\{\nabla\phi\}(\z),\label{js_4well} 
\end{align}
is an irreversible drift that is by construction orthogonal to
$\nabla\phi$ and thus does not alter the steady-state (i.e.\ same
$\ps=\exp[-\phi]$ for equilibrium ($\Omega=0$) or any other $\Omega$). With
Eq.~\eqref{js_4well} the dissipation in Eq.~\eqref{EPR} for this process reads
\begin{align}
&\dot{\Sigma}=D\Omega^2\int{\rm d}^2\mathbf{x}\{\nabla\phi\}(\mathbf x)^T\begin{bmatrix}0&-1\\1&0\end{bmatrix}^T\begin{bmatrix}0&-1\\1&0\end{bmatrix}\cdot\\
&\{\nabla\phi\}(\mathbf x)p_s(\mathbf{x})=D\Omega^2\int{\rm d}^2\mathbf{x}\{\nabla\phi\}^2(\mathbf x)\exp\left[-\phi(\mathbf x)\right],\nonumber
\end{align}
which is solved numerically and gives $\dot{\Sigma}=19.65 D\Omega^2$. We see in Fig.~\ref{fig:TUR}h that some intermediate coarse graining $h$ is still optimal, but the optimal scale $h$ now
depends more intricately on $\x$ and the curves are not convex in $h$
anymore. 

Overall we see that the approach is robust and easily applicable, and
does not require to determine and parameterize any rates. Moreover, due to the implications of the 
theory to the limits $h\to 0,\infty$ we can assert that some intermediate coarse graining will generally be optimal.

\section{Simplifications and symmetries}\label{sec:symmetry}

In this section we list the symmetries obeyed by the results in
Eqs.~\eqref{mean_rho},\eqref{mean_j_part2},\eqref{density-density
  result},\eqref{current-density-result},\eqref{current covariance
  result} (with integral operator \eqref{int_op_simpl}).
Note that the limit $h\to 0$ was carried out 
\blue{in Sec.~\ref{sec:h to 0}} 
and the limit $h\to\infty$ gives $U^h_\x=c$
as noted before 
which greatly simplifies the further analysis. The limits $t\to 0$ and
$t\to\infty$ will be addressed in Section \ref{short and long times}
(see also Supplemental Material in \cite{AccompanyingLetter}).

First consider dynamics obeying detailed balance, i.e.\ $\js=\f 0$. We then
have $\revj_\z=-\bj_\z=-\bj^g(\z)$ and the dual-reversal symmetry in
Eq.~\eqref{dual_rev}, simplifies to the detailed balance statement
$G(\y,t|\x)\ps(\x)=G(\x,t|\y)\ps(\y)$ or
$P_\z(\z',t)=P_{\z'}(\z,t)$. From this we obtain the following
simplifications for $\js=\f 0$:
\begin{align}
&\Es{\overline{\f J^U_\x}(t)}=\f 0,\quad \f C^{\x\y}_{\f J \rho}(t)=\f 0\nonumber\\
&C^{\x\y}_{\rho\rho}(t)=2\integrals^{t,U}_{\x\y}\left[P_\z(\z',t')-\ps(\z)\ps(\z')\right],\nonumber\\
&C^{\x\y}_{\f J\cdot \f J}(t)=\frac2t\int d\z {\rm Tr}[\f D(\z)]U^h_\x(\z)U^h_\y(\z)\ps(\z)\nonumber\\
&-2\integrals_{\x\y}^{t,U}[\bj^g(\z)\cdot\bj^g(\z')P_\z(\z',t')+\js(\z)\cdot\js(\z')].\label{results in detailed balance}
\end{align}
For the remainder of this section we consider $\js\neq\f 0$. Note that
by definition  the interchange $\x\leftrightarrow\y$ leaves
$C^{\x\y}_{\rho\rho}(t)$ and $C^{\x\y}_{\f J\cdot \f J}(t)$ invariant,
but \emph{not} $C^{\x\y}_{\f J\rho}(t)$ since it considers currents at $\x$ and densities at $\y$.

For single-point correlations and variances $\x=\y$ (more precisely
$U^h_\x=U^h_\y$) the integrations over $\z$ and $\z'$ are equivalent
and thus the results simplify to 
\begin{align}
&C^{\x\x}_{\rho\rho}(t)=2\integrals^{t,U}_{\x\x}\left[P_\z(\z',t')-\ps(\z)\ps(\z')\right]\nonumber\\
&\f C^{\x\x}_{\f J \rho}(t)=\integrals_{\x\x}^{t,U}[(\bj_{\z'}+\revj_\z)P_\z(\z',t')-2\js(\z)\ps(\z')]\nonumber\\
&C^{\x\x}_{\f J\cdot \f J}(t)=\frac2t\int d\z {\rm Tr}[\f D(\z)][U^h_\x]^2(\z)\ps(\z)\nonumber\\
&+2\integrals_{\x\x}^{t,U}[\revj_\z\cdot\bj_{\z'}P_\z(\z',t')-\js(\z)\cdot\js(\z')].\label{results for x=y}
\end{align}
Now we again allow $\x\neq\y$ and consider the process and the
$\js\leftrightarrow-\js$ inverted process.  Then, from
Eq.~\eqref{dual_rev} and $[\revj_{\z'}\cdot\bj_\z]^{-\js}=-\bj_{\z'}\cdot[-\revj_\z]=\bj_{\z'}\cdot\revj_\z=\revj_\z\cdot\bj_{\z'}$, we get $[\revj_{\z'}\cdot\bj_\z P_{\z'}(\z,t')]^{-\js}=\revj_\z\cdot\bj_{\z'} P_\z(\z',t')$ and thus obtain
\begin{align}
\Es{\overline{\rho^U_\x}(t)}&=\Es{\overline{\rho^U_\x}(t)}^{-\js}\nonumber\\
\Es{\overline{\f J^U_\x}(t)}&=-\Es{\overline{\f J^U_\x}(t)}^{-\js}\nonumber\\
C^{\x\y}_{\rho\rho}(t)&=[C^{\x\y}_{\rho\rho}(t)]^{-\js}\nonumber\\
\f C^{\x\y}_{\f J \rho}(t)&=-[\f C^{\x\y}_{\f J \rho}(t)]^{-\js}\nonumber\\
C^{\x\y}_{\f J\cdot \f J}(t)&=[C^{\x\y}_{\f J\cdot \f J}(t)]^{-\js}.
\label{results symmetries for -js}
\end{align}
In addition to the symmetries of the first and second
cumulants, a stronger path-wise version of the dual-reversal symmetry
in Eq.~\eqref{dual_rev} (or time-reversal symmetry at equilibrium)
dictates symmetries of the full distributions of the functionals of
steady-state trajectories under the reversal $\js\leftrightarrow-\js$.
Notably, at equilibrium ($\js=\f 0$) these simplify to symmetries of
the process (which is a much stronger result since we do not have to
compare to another (artificial) process with an inverted $\js$). 

To motivate this stronger symmetry, note that for steady-state initial conditions for any finite set of times
$t_1<t_2<\dots<t_n$ we have that the joint density $P_n(\dots)$ for
positions $z_i$ at equally spaced times $t_i=i\times\Delta t$ for
$i=0,1,\dots,n$ is given by (since we have a Markov process by
definition, i.e. Eq.~\eqref{SDE} has no memory)
\begin{align}
&P_n(z_0,t_0;z_1,t_1;\dots;z_n,t_n)\nonumber\\
&=\ps(\x_0)G(\x_1,\Delta t|\x_0)\cdots G(\x_n,\Delta t|\x_{n-1}).
\end{align}
By applying the dual-reversal symmetry Eq.~\eqref{dual_rev} $n-1$ times, we obtain
\begin{align}
&P_n(z_0,t_0;z_1,t_1;\dots;z_n,t_n)\nonumber\\
&=G^{-\js}(\x_0,\Delta t|\x_1)\cdots G^{-\js}(\x_{n-1},\Delta t|\x_n)\ps(\x_n)\nonumber\\
&=P^{-\js}_n(z_n,0;z_{n-1},\Delta t;\dots;z_0,n\Delta t)\nonumber\\
&=P^{-\js}_n(z_n,t_0;z_{n-1},t_1;\dots;z_0,t_n).\label{discrete time path reversed} 
\end{align}
The $n+1$ points $(z_1,\dots,z_n)$ represent a discrete-time path for which Eq.~\eqref{discrete time path reversed} implies the path-wise discrete-time dual-reversal symmetry (denote $t=t_n=n\Delta t$)
\begin{align}
&P_n(z_0,t_0;z_1,t_1;\dots;z_n,t_n)\nonumber\\
&=P^{-\js}_n(z_n,t-t_n;z_{n-1},t-t_{n-1};\dots;z_0,t-t_0),\label{dual reversal discrete path}
\end{align}
i.e.\ the probability of forward paths $(\x_{t_i})_{i=0,1,\dots,n}$ agrees with the probability of backwards paths of the process with inverted steady-state current $\js\to-\js$, i.e. 
\begin{align}
\mathbb P\left[(\x_{t_i})_{i=0,1,\dots,n}\right]=\mathbb P^{-\js}\left[(\x_{t-t_i})_{i=0,1,\dots,n}\right].\label{dual reversal discrete path 2}
\end{align}
Note that at equilibrium, $\js=\f 0$, this is nothing but  the
detailed balance for discrete-time paths. 

Assuming that one can take a continuum limit $\Delta t\to 0$ (and that
a resulting path measure exits) one could conclude that continuous time paths fulfill the symmetry (see also \cite{Dechant2021PRR})
\begin{align}
\mathbb P\left[(\x_\tau)_{0\le\tau\le t}\right]=\mathbb P^{-\js}\left[(\x_{t-\tau})_{0\le\tau\le t}\right].\label{dual reversal path}
\end{align}
Based on this strong symmetry, and noting that densities are symmetric while currents are antisymmetric under time reversal, i.e.\ 
\begin{align}
\overline{\rho^U_\x}\left[(x_\tau)_{0\le\tau\le t}\right]&=\overline{\rho^U_\x}\left[(x_{t-\tau})_{0\le\tau\le t}\right]\nonumber\\
\overline{\f J^U_\x}\left[(x_\tau)_{0\le\tau\le t}\right]&=-\overline{\f J^U_\x}\left[(x_{t-\tau})_{0\le\tau\le t}\right],
\end{align}
we obtain the following symmetries
\begin{align}
\mathbb P\left[\overline{\rho^U_\x}(t)=u\right]&=\mathbb P^{-\js}\left[\overline{\rho^U_\x}(t)=u\right]\nonumber\\
\mathbb P\left[\overline{\f J^U_\x}(t)=\f u\right]&=\mathbb P^{-\js}\left[\overline{\f J^U_\x}(t)=-\f u\right].
\label{full symmetries for -js}
\end{align}
Eq.~\eqref{full symmetries for -js} implies symmetries for mean values and variances ($\x=\y$) listed in Eq.~\eqref{results symmetries for -js} since it implies that all moments of $\overline{\rho^U_\x}(t)$ agree and that the $n$-th moment of a current component $i$ fulfills $\Es{{[\overline{\f J^U_\x}(t)]_i}^{\!n}}=\Es{{[-\overline{\f J^U_\x}(t)]_i}^{\!n}}^{-\js}=(-1)^n\Es{{[\overline{\f J^U_\x}(t)]_i}^{\!n}}^{-\js}$.

Note that Eq.~\eqref{full symmetries for -js} implies that the
statistics of $\rho(t)$ (incl.\ all moments) in general depends on
$\js$ but is invariant under the inversion
$\js\leftrightarrow-\js$. Moreover,  current fluctuations at equilibrium ($\js=\f 0$, hence $\mathbb P_{\rm EQ}\equiv\mathbb P=\mathbb P^{-\js}$) are symmetric around the mean $\Es{\overline{\f J^U_\x}}=\f 0$, i.e.\
\begin{align}
\mathbb P_{\rm EQ}\left[\overline{\f J^U_\x}(t)=\f u\right]&=\mathbb P_{\rm EQ}\left[\overline{\f J^U_\x}(t)=-\f u\right].
\end{align}
The symmetries for correlations in Eq.~\eqref{results symmetries for -js}, possibly with $\x\ne\y$, may be seen as implications of the more general symmetries
\begin{align}
\mathbb P\left[\overline{\rho^U_\x}(t)\overline{\rho^U_\y}(t)=u\right]&=\mathbb P^{-\js}\left[\overline{\rho^U_\x}(t)\overline{\rho^U_\y}(t)=u\right]\nonumber\\
\mathbb P\left[\overline{\f J^U_\x}(t)\overline{\rho^U_\y}(t)=\f u\right]&=\mathbb P^{-\js}\left[\overline{\f J^U_\x}(t)\overline{\rho^U_\y}(t)=\f u\right]\nonumber\\
\mathbb P\left[\overline{\f J^U_\x}(t)\cdot\overline{\f J^U_\y}(t)=u\right]&=\mathbb P^{-\js}\left[\overline{\f J^U_\x}(t)\cdot\overline{\f J^U_\y}(t)=u\right].
\label{full correlation symmetries for -js}
\end{align}

\blue{\section{Continuity equation along individual diffusion paths}\label{cont eq} 
In this section we derive a continuity equation for the
time-accumulated density $t\overline{\rho^U_\x}(t)$ and current
$t\overline{\f J^U_\x}(t)$ defined with windows that satisfy
$U^h_\x(\x')=U^h_{\f 0}(\x'-\x)$.  This condition in particular holds
for all window functions that may be interpreted as a spatial coarse
graining, as e.g.\ a Gaussian around $\x$ or any indicator function $U^h_\x(\x')\propto\mathbbm1_{\norm{\x'-\x}\le h}$ with any norm $\norm\cdots$. Under this assumption, $-\nabla_\x U^h_\x(\x')=\nabla_{\x'} U^h_\x(\x')\equiv \left\{\nabla U^h_\x\right \}(\x')$ such that 
\begin{align}
&-\nabla_{\x}\int_{\tau=0}^{\tau=t}U^h_{\x}(\x_\tau)\circ\rmd\x_\tau
=\int_{\tau=0}^{\tau=t}\left\{\nabla U^h_{\x}\right \}(\x_\tau)\circ\rmd\x_\tau\nonumber\\
&\qquad=U^h_\x(\x_t)-U^h_\x(\x_0)
=\partial_t\int_0^t U^h_{\x}(\x_\tau)\rmd\tau,
\end{align}
which can be written in the form of a continuity equation
\begin{equation}
\partial_t[t\overline{\rho^U_\x}(t)]=-\nabla_{\x}\cdot t\,\overline{\f J^U_\x}(t).\label{continuity_eq} 
\end{equation}
This generalizes the notion of a continuity equation to individual
trajectories $(\x_\tau)_{0\le\tau\le t}$ with arbitrary initial and
end points. For steady-state dynamics and normalized window functions,
i.e.\ $\int\rmd^d z U^h_\x(\z)=1$, taking the mean $\E{\cdot}_{\rm s}$
of Eq.~\eqref{continuity_eq} leads to a continuity equation for
(coarse-grained) probability densities. Conversely, for non-normalized
window functions $\int\rmd^d z U^h_\x(\z)={\rm Volume}(U^h_\x)$, the
mean $\E{\cdot}_{\rm s}$ of Eq.~\eqref{continuity_eq} may be
interpreted as a continuity equation for probabilities. 

Note that 
the statement
$\int_{\tau=0}^{\tau=t}\left\{\nabla
U^h_{\x}\right\}(\x_\tau)\circ\rmd\x_\tau=U^h_\x(\x_t)-U^h_\x(\x_0)$
holds only for the Stratonovich integral but, e.g.,\ not for an It\^o
integral. Therefore, the continuity equation further motivates the
definition Eq.~\eqref{def_current} via the Stratonovich integral,
which was also required for the mean empirical current (see comment
below Eq.~\eqref{mean_j_part2}), and for consistency of time reversal
(e.g.\ to obtain the symmetry in Eqs.~\eqref{current-density-result}
and \eqref{current covariance result}; also see Fig.~2 in
\cite{AccompanyingLetter}). 
}

\section{Short and long trajectories and the central-limit regime}\label{short and long times} \begin{figure*}
  \centering
  \includegraphics[width=0.85\textwidth]{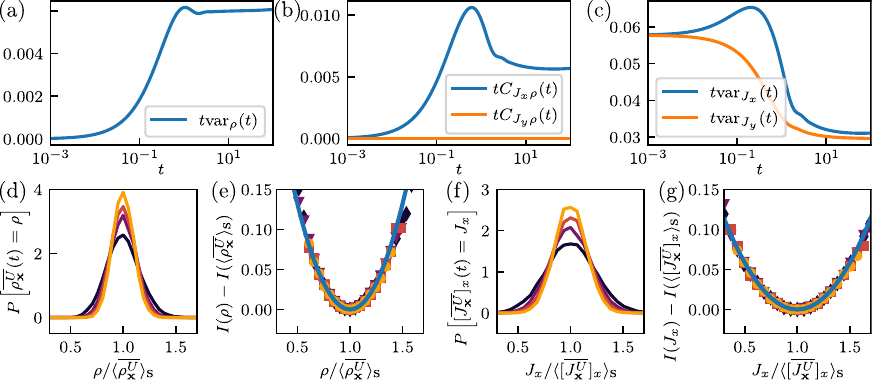}\caption{
We consider the rotational flow Eq.~\eqref{OUP} with $\Omega=3$ starting from steady-state
initial conditions and use a Gaussian coarse-graining window Eq.~\eqref{Gauss window} around $\x=(0,1)^T$ with width $h=0.5$. (a) Analytical result for the variance of the time-averaged density $\overline{\rho^U_\x(t)}$ multiplied by time $t$
as a function of $t$. At long times the variance
approaches the large deviation variance in Eq.~\eqref{large deviation
  variance}. (b) As in (a) but for the components of the correlation
vector $\f C^{\x\x}_{\f J \rho}(t)$ as in Eqs.~\eqref{current-density-result} and \eqref{results for x=y}. (c) As
in (a) but for the variances of the current components Eq.~\eqref{current
  component variance result}. 
(d) Simulation of the probability density function of the empirical
density $\overline{\rho^U_\x}(t)$ assuming the parameters listed
above. Colors of lines and symbols throughout denote $t=40,60,80,100$
from dark to bright. The simulated probability densities were obtained
from histograms of $2\times 10^{4}$ trajectories for each set of
parameters.  (e) Parabolic \finalchange{approximation for the} rate function with variance from 
Eq.~\eqref{large deviation variance} (line) and simulated rate function $I(\rho)=-\frac{1}{t}\ln P[\overline{\rho^U_\x}(t)=\rho]$ (symbols). The numerical value of the rate function at the mean $\rho=\langle\overline{\rho^U_\x}(t)\rangle_{\rm s}$ was subtracted. 
(f-g) As in (d-e) but for the $x$-component of the current
$[\overline{\f J^U_\x}]_x$ instead of the density
$\overline{\rho^U_\x}$.}
\label{fig:towards_LD_and_LDT}
\end{figure*}

As already noted on several occasions, in the case of  steady-state
initial conditions the mean values of the time-averaged density and
current are time-independent, see
Eqs.~\eqref{mean_rho},\eqref{mean_j_part2}. The correlation and
(co)variance results (Eqs.~\eqref{density-density
  result},\eqref{current-density-result},\eqref{current covariance
  result} with integral operator \eqref{int_op_simpl}) display a non-trivial temporal behavior dictated by the time integrals $\frac{1}{t}\!\int_0^t\rmd t'\left(1-\frac{t'}{t}\right)$ over two-point densities $P_\z(\z',t')$.

In Fig.~\ref{fig:towards_LD_and_LDT}a-c we depict this time-dependent
behavior for the two-dimensional rotational flow Eq.~\eqref{OUP} for $\x=\y$. The
short-time behavior can be obtained by analogy to the short-time
expansion in the SM of \cite{AccompanyingLetter}. \blue{Note that the
  short-time limit of fluctuations of time-integrated currents
  recently attracted much attention in the context of inference of
  dissipation, since in this limit the thermodynamic uncertainty relation
  becomes sharp \cite{Supriya,Otsubo2020PRE}.}
The long-time
behavior shows that $\f C(t),{\rm var}(t)\propto t^{-1}$, as expected
from the central limit theorem (and large deviation theory) due to sufficiently many
sufficiently uncorrelated visits of the window region. Accordingly, a
serious problem is encountered in dimensions $\ge 2$ in the limit
$h\to 0$ because diffusive trajectories do not hit points (for a
detailed discussion see \cite{AccompanyingLetter}).

The limit of $t\f C(t),\, t{\rm var}(t)$ for large $t$ can be obtained
as follows. We have $\int_{t'}^\infty\rmd t''[P_\y(\x,t'')-\ps(\x)]\to
0$ for $t'\to\infty$ since
$P_\y(\x,t')\overset{t'\to\infty}\longrightarrow\ps(\x)$ and $\bj_\x
P_\y(\x,t')\overset{t'\to\infty}\longrightarrow\js(\x)$ with
exponentially decaying deviations.  This implies that for large $t$,
we can replace $\frac{1}{t}\int_0^t\rmd t'\left (1-\frac{t'}{t}\right
)$ by $\frac{1}{t}\int_0^\infty\rmd t'$ in the integral operator
\eqref{int_op_simpl}. This replacement of integrals
and the scaling are also confirmed by a spectral expansion (see
e.g.\ \cite{Lapolla2020PRR} for spectral-theoretic results for the
empirical density). 

We now \finalchange{discuss the central-limit regime, which is contained in large deviation theory as small deviations from the mean}.  
According to the central limit theorem \blue{(for not almost surely constant $U_\x^h$, and for finite variances (i.e.\ strictly positive $h$, see \ref{sec:h to 0}))}, the probability distributions $p(A_t=a)$ 
for $A_t=\overline{\rho^U_\x}(t)$ and $A_t=\overline{\f J^U_\x}(t)$ become Gaussian for large $t$\finalchange{. This is contained in large deviation theory in terms of a parabola that locally (for $a\approx\mu$) approximates the rate function}
\begin{align}
I(a)=-\lim_{t\to\infty}\frac{1}{t}\ln p(A_t=a)\finalchange{\approx}\frac{(a-\mu)^2}{2\sigma^2_A},\label{rate function Gaussian} 
\end{align}
where the mean $\mu$ is given by $\E{\rho^U_\x(t)}_{\rm
  s}=\int\rmd^dzU^h_\x(\z)\ps(\z)$ and $\E{\f J^U_\x(t)}_{\rm
  s}=\int\rmd^dzU^h_\x(\z)\js(\z)$ (see
Eqs.~\eqref{mean_rho},\eqref{mean_j_part2}) and the large deviation
variance $\sigma^2_A$ follows by the above arguments from
Eqs.~\eqref{density-density result} and \eqref{current covariance
  result} for $\x=\y$ as in Eq.~\eqref{results for x=y} as
\begin{align}
\sigma^2_{\overline{\rho^U_\x}}&\equiv\lim_{t\to\infty}t\,{\rm var}_\rho^\x(t)\nonumber\\
&= 2\!\int_0^\infty\!\!\!\!\!\!\rmd t'\!\!\int\!\!\rmd^dz\!\int\!\!\rmd^dz'
  U^h_\x(\z)U^h_\x(\z')\nonumber\\
  &\quad\times \left[P_{\z'}(\z,t)-\ps(\z)\ps(\z')\right ],\label{large deviation variance density}
\end{align}
as well as
\begin{align}
&\sigma^2_{\overline{\f J^U_\x}}\equiv\lim_{t\to\infty}t\,{\rm var}_\f J^\x(t)=2{\rm Tr}\f D\int\rmd^d z\, [U^h_\x]^2(\z)\ps(\z)\nonumber\\
&+2\int_0^\infty\!\!\!\!\!\!\rmd t'\!\!\int\!\!\rmd^dz\!\int\!\!\rmd^dz'
  U^h_\x(\z)U^h_\x(\z')\nonumber\\
  &\quad \times \left [\bj_{\z}\cdot\revj_{\z'}P_{\z'}(\z,t)-\js(\z)\js(\z')\right ].\label{large deviation variance} 
\end{align}
For any Lebesgue integrable window function $U^h_\x$ (i.e.\ if the window
size $h$ fulfills $h>0$), and in $d=1$ even for the delta-function, this variance is finite, and
the central limit theorem applies as described above. The parabolic
\finalchange{approximation for the} rate function for a two dimensional system with finite window size
$h>0$ is shown for the density $\overline{\rho^U_\x}(t)$ and current
$\overline{\f J^U_\x}(t)$ in Fig.~\ref{fig:towards_LD_and_LDT}e and
g. The agreement of the simulation and the variance given by Eqs.~\eqref{large deviation variance density}-\eqref{large deviation variance} is readily confirmed.

\blue{If we instead take the limit of no coarse graining $h\to 0$ in
  multi-dimensional space $d\ge 2$, the variances diverge (see
  Eq.~\eqref{bounds h to 0}). Fig.~\ref{FgLD2} depicts the
  distribution of the 
empirical density $\overline{\rho^U_\x}(t)$ in a fixed point $\x$ for
different times $t$ and window sizes $h$. We see that the distribution
becomes non-Gaussian for small $h$, in particular the most probable
value departs from the mean and approaches $0$. Even though a Gaussian
distribution is restored for longer times (see  Fig.~\ref{FgLD2}b),
for even smaller window sizes the distribution again becomes
non-Gaussian (see  Fig.~\ref{FgLD2}c). This behavior is not
  surprising since Gaussian distributions are only expected for
  sufficiently many
  (sufficiently uncorrelated) visits of the coarse graining
  window. For $h\to 0$ the recurrence time to return to the window
  diverges and thus for any finite $t$ one cannot expect a Gaussian
  distribution. Note that it is not clear whether a limit in
  distribution for $h\to 0$ of $\overline{\rho^U_\x}$ and
  $\overline{\f J^U_\x}$ even exists. We hypothesize that, if the a
  limit $h\to 0$ of the distribution indeed exists, then it does so only as a scaling limit with $h\to 0$ and $t\to\infty$ simultaneously.

\begin{figure}[!ht]
\begin{center}
\includegraphics[width=.48\textwidth]{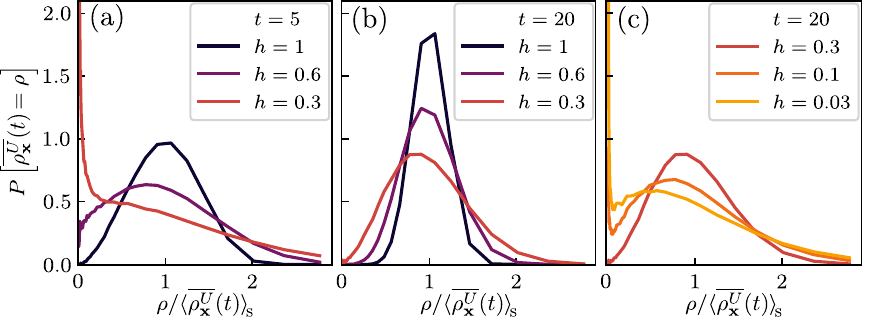}
\caption{\label{FgLD2}
\blue{Simulation of the probability density function of the empirical density $\overline{\rho^U_\x}(t)$ for $\x=(0,1)^T$ and Gaussian window function $U^h_\x$ Eq.~\eqref{Gauss window} with width $h$ for the two-dimensional driven Ornstein-Uhlenbeck process Eq.~\eqref{OUP} with $\Omega=3$ with $\x_0$ starting from the steady state. The simulated probability density were obtained from histograms of $2\times 10^{5}$ trajectories for each set of parameters.}}
\end{center}\end{figure}}

\blue{
\section{Outlook beyond overdamped dynamics}\label{sec:outlook} 
In this section we give a brief outlook on the relevance of our
findings in the limit $h\to 0$ for processes that are \emph{not} described by
purely overdamped dynamics. In particular, we highlight that although
in physical systems the assumption of overdamped dynamics breaks down
at very small time or length scales (which often may not be
observable), the predicted divergence of fluctuations in the limit $h\to 0$ does
not break down, or at least it remains true for sufficiently small finite
$h$ that 
empirical densities and currents 
attain numerically very large values, i.e.\ effectively diverge. We emphasize
that this section only establishes an outlook that underscores the
experimental relevance of our approach, but does not contain
quantitative theoretical results. 
Note that beyond the examples given here, the results in the limit
$h\to 0$ also apply to Markov jump processes as illustrated in the
Supplemental Material of \cite{AccompanyingLetter}.

To go beyond the assumption of Markovian overdamped motion assumed in
Eq.~\eqref{SDE}, Fig.~\ref{fig:experiment_underdamped} depicts the
fluctuations of the empirical density and current for two very
different types of stochastic dynamics.  
\begin{figure}[ht!!]
  \centering
  \includegraphics[width=.45\textwidth]{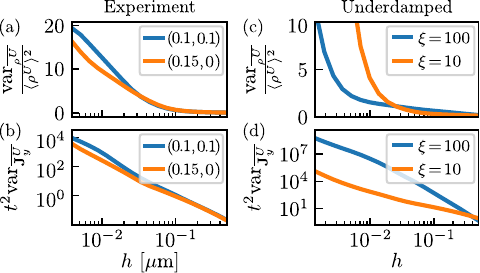}
  \caption{\blue{(a) Variance divided by squared mean of the empirical
      density and (b) variance of the empirical current with Gaussian
      coarse graining window (see Eqs~\eqref{def_current},\eqref{Gauss
        window}) around $\x=(0.1,0.1)^T\mu$m and $\x=(0.15,0)^T\mu$m
      for $200$ experimental trajectories $(\x_\tau-\x_0)_{\tau\le
        50\ \rm s}$ measured in particle tracking in living cells
      \cite{Sabri2020PRL} with time-step $\rmd t=0.1$ s and spatial
      resolution $10^{-3}\ \mu$m. (c) and (d) depict analogous results for
      underdamped dynamics simulated according to
      Eq.~\eqref{underdamped} for $1000$ trajectories with an Euler
      integration scheme with time step $\rmd t=0.02$, total time $10$,    
       initial positions $\x_0=(0,0)^T$ and Maxwellian initial velocities $\f v_0$ (i.e.\ zero-mean Gaussian with variance $k_BT$),
      evaluated with Gaussian
      coarse graining window around $\x=(0.1,0)^T$. The observables in a,c and b,d
      relate to the predicted analytical curves in 
      Figs.~\ref{fig:rel_err_rho} and \ref{fig:TUR}c, respectively,
      but now for completely different underlying dynamics. We note that
      processes not following an overdamped motion, in fact even
      non-Markovian processes, also seem to
      feature the divergence of fluctuations in the limit $h\to 0$, at least down to relevant
      small values of $h>0$ (larger than the resolution yet small
      enough to display impractically large fluctuations to be experimentally useful).} 
    \label{fig:experiment_underdamped}}
\end{figure} 
In Fig.~\ref{fig:experiment_underdamped}a,b  we 
evaluate the functionals in Eq.~\eqref{def_current} with a Gaussian window
function from Eq.~\eqref{Gauss window} for particle-tracking data in living
cells that was found to be well described by a two-state fractional Brownian
motion \cite{Sabri2020PRL,Dieball2022NJP,Janczura2021NJP}. The latter in
particular is non-Markovian with subdiffusive anti-persistence on
a given time scale. 
We observe that, even though the assumption of Markovian overdamped motion
Eq.~\eqref{SDE} is obviously violated (on some time and spatial scales), and
thus the results of our work do not necessarily apply, we still find
divergent fluctuations in the limit of small coarse graining scales
$h$.  Note that the resolution of the measurement is $h=10^{-3}\ \mu$m
\cite{Sabri2020PRL}. In
Fig.~\ref{fig:experiment_underdamped}a,b we observe that even for $h$
above this resolution limit the fluctuations approach impracticably
large values.  Therefore, we propose that in general scenarios
(e.g.\ in this experimental set-up that extends way beyond the discussed
overdamped dynamics) coarse graining empirical densities and currents
may even in the case of very good statistics be necessary to obtain
experimentally meaningful values with limited fluctuations. 

In Fig.~\ref{fig:experiment_underdamped}c,d we similarly evaluate the
coarse-grained empirical density and current for two-dimensional
\emph{underdamped} harmonically confined Langevin dynamics with friction constant $\xi$ (setting
for convenience the mass $m=1$ and temperature $k_BT=1$) simulated by
integrating the equations of motion
\begin{align}
\rmd \x_t&=\f v_t\rmd t\nonumber\\
\rmd \f v_t&=-\xi \f v_t\rmd t-\f x_t\rmd t+\sqrt{2\xi}\rmd\f W_t.\label{underdamped} 
\end{align}
This dynamics exhibits persistence on time scales around or below
$m/\xi$  (i.e.\ the ballistic regime). Again we find in
Fig.~\ref{fig:experiment_underdamped}c,d  that the divergence
predicted in the limit $h\to 0$ for overdamped dynamics is
qualitatively preserved. The quantitative order of divergence will
depend on the details of the process. We hypothesize that on time
scales $h^2/D$ (with diffusion constant $D=k_BT/\xi$) that are smaller
that $m/\xi$ the ballistic regime will cause deviations from the predicted
divergence results in \cite{AccompanyingLetter}. Following the
arguments in Sec.~\ref{sec:h to 0}, the expressions will still
diverge since the probability to hit points in $d\ge 2$-dimensional
space becomes zero.
 
The influence of the details of the process, such as memory effects
and ballistic transport, constitutes an interesting direction for
future research that, however, goes beyond the scope of the present
work. From the qualitative behavior found in
Fig.~\ref{fig:experiment_underdamped} we may already conclude  that
the relevance of coarse graining empirical densities and currents to
ensure finite and manageable fluctuations appears to be a quite general
result, exceeding beyond the overdamped dynamics discussed in this
work.  
}

\section{Conclusion}
In this extended expos\'e accompanying the Letter
\cite{AccompanyingLetter} we presented the conceptual and technical
background that is required to describe and understand the statistics
of the empirical density and current of steady-state diffusions, which are central to 
statistical mechanics and thermodynamics on the level of individual
trajectories. In order to gain deeper insight into the meaning of
fluctuations of the empirical density and current we made use of a generalized
time-reversal symmetry. 
We carried out a systematic analysis of the effect of a
spatial coarse graining. A systematic variation of the
coarse-graining scale in an \emph{a posteriori} smoothing of
trajectory data was proposed as an efficient method to infer bounds on
a system's dissipation. Moreover, we discussed symmetries in the
statistics of the empirical current and density that arise as a result
of the (generalized) time-reversal symmetry. 
Throughout the work we advocated the application of stochastic calculus, which
is very powerful in the analysis of related problems and represents a
more direct alternative to Feynman-Kac theory and path-integral
methods. The technical background and concepts presented here may serve as a basis for forthcoming publications, including the generalization of the presented inference strategy to windows that are not centered at an individual point, as well as the use of the correlations result entering the CTUR inequality \cite{Dechant2021PRX}.

\emph{Acknowledgments---}\blue{We thank Diego Krapf and
  Matthias Weiss for kindly providing traces from their particle-tracking experiments.} Financial support from Studienstiftung des Deutschen Volkes (to
C.\ D.) and the German Research Foundation (DFG) through the Emmy
Noether Program GO 2762/1-2 (to A.\ G.)  is gratefully acknowledged.

\onecolumngrid
\appendix
\blue{\section{Derivations in the limit of no coarse graining}\label{sec:app}
We now take the limit to very small window sizes, i.e.\ the limit to no coarse graining, which will turn out
to depend only on the properties of the two-point functions
$P_\x(\y,t')$ for small time differences $t'=t_2-t_1$. This allows us
to derive the bounds in Eq.~\eqref{bounds h to 0}. We consider
normalized window functions such that for a window size $h\to 0$ the window function becomes a delta distribution $U^h_\x(\z)\to\delta(\x-\z)$.

\subsection{Density variance}
For the variance of the density ${\rm var}^\x_\rho(t)\equiv\E{\overline{\rho^U_\x}(t)^2}_{\rm s}-\E{\overline{\rho^U_\x}(t)}_{\rm s}^2$, we have (see Eq.~\eqref{density-density result})
\begin{align}
{\rm var}^\x_\rho(t)&=2\integrals^{t,U}_{\x\x}[P_{\z'}(\z,t')-\ps(\z)\ps(\z')].
\end{align}
For window size $h\to 0$ the mean remains finite such that $\integrals^{t,U}_{\x\x}[\ps(\z)\ps(\z')]\overset{h\to0}\longrightarrow-2\ps(\x)^2=O(h^0)$. Now consider
\begin{align}
\integrals^{t,U}_{\x\x}[P_{\z'}(\z,t')]=\frac{1}{t}\int_0^t\rmd t'\left (1-\frac{t'}{t}\right )\int\rmd^dz\int\rmd^dz' U^h_\x(\z)U^h_\x(\z')P_{\z'}(\z,t').
\end{align}
For $t'>\varepsilon>0$, $P_{\z'}(\z,t')$ is bounded and thus
$\int\rmd^dz\int\rmd^dz' U^h_\x(\z)U^h_\x(\z')P_{\z'}(\z,t')$ is bounded
using $\norm{P_{\z'}(\z,\varepsilon)}_\infty=O(h^0)$. Contributions
diverging for $h\to 0$ can thus only come from the $t'\to 0$ part of
the integral, i.e.\ from \textbf{small time differences} $t'=t_2-t_1$
(but \textbf{not} small absolute time $t$) in the Dyson series. To get the dominant divergent contribution, we can thus set $1-t'/t\to 1$ and replace the two-point function $P_{\z'}(\z,t')$ by the short time propagator $P_\y(\z,t')\to\ps(\z')G_{\rm short}(\z,t'|\z')$ which reads (for simplicity take $\f D(\z)=D\f 1$, which we generalize later) \cite{Risken1996} 
\begin{align}
G_{\rm short}(\z,t'|\z')&=(4\pi Dt')^{-d/2}\exp\left[-\frac{\left [\z-\z'-F(\z')t'\right]^2}{4Dt'}\right]\nonumber\\
&=(4\pi Dt')^{-d/2}\exp\left[-\frac{\left [\z-\z'\right]^2}{4Dt'}+\frac{2(\z-\z')\cdot\f F(\z')t'}{4Dt'}+O(t')\right]\nonumber\\
&\overset{\z\approx\z'}\approx(4\pi Dt')^{-d/2}\left[1+\frac{1}{4D}(\z-\z')\cdot\f F(\z')\right]\exp\left[-\frac{\left [\z-\z'\right]^2}{4Dt'}\right].
\end{align}
We write for $t'\to 0, \z-\z'\to\f 0$,
\begin{align}
G_{\rm short,2}&\equiv(4\pi Dt')^{-d/2}\left[1+\frac{1}{2D}(\z-\z')\cdot\f F(\z')\right]\exp\left[-\frac{\left [\z-\z'\right]^2}{4Dt'}\right]\nonumber\\
G_{\rm short,3}&\equiv(4\pi Dt')^{-d/2}\exp\left[-\frac{\left [\z-\z'\right]^2}{4Dt'}\right],
\label{short time propagators} \end{align}
where $G_{\rm short,2}$ can be replaced by $G_{\rm short,3}$ (since $\z-\z'$ is small) if $G_{\rm short,3}$ does not give zero in the integrals..

For Gaussian window functions
\begin{align}
U^h_\x(\z)=(2\pi h^2)^{-d/2}\exp\left[-\frac{(\z-\x)^2}{2h^2}\right], 
\end{align}
we obtain for the spatial integrals
\begin{align}
\int\rmd^dz\int\rmd^dz' & U^h_\x(\z)U^h_\x(\z')G_{\rm short,3}(\z,t'|\z')\ps(\z')\nonumber\\
&\simeq
\ps(\x)\int\rmd^dz\int\rmd^dz' U^h_\x(\z)U^h_\x(\z')G_{\rm short,3}(\z,t'|\z')\nonumber\\
&=\ps(\x)(2\pi h^2)^{-d}(4\pi Dt')^{-d/2}\int\rmd^dz\int\rmd^dz'\exp\left [-\frac{(\z-\z)^2}{2h^2}-\frac{(\z'-\z)^2}{2h^2}-\frac{(\z-\z')^2}{4Dt'}\right]\nonumber\\
&=\ps(\x)(2\pi h^2)^{-d}(4\pi Dt')^{-d/2}\left(\int\rmd x_1\int\rmd y_1\exp\left [-\frac{x_1^2}{2h^2}-\frac{y_1^2}{2h^2}-\frac{(x_1-y_1)^2}{4Dt'}\right]\right )^d\nonumber\\
&=\ps(\x)\left[\frac{\sqrt{2 D h^{2} t' + h^{4}}}{2 \sqrt{\pi} h^{2} \sqrt{2 D t' + h^{2}} \sqrt{\frac{D t'}{h^{2}} + 1}}\right]^d\nonumber\\
&=\ps(\x)(4\pi)^{-d/2}(Dt'+h^2)^{-d/2}.
\end{align}
This implies, throughout denoting by $\simeq$ asymptotic equality in
the limit $h\to 0$ (i.e.\ equality of the largest order),
\begin{align}
\integrals^{t,U}_{\x\x}[P_{\z'}(\z,t')]&\simeq(4\pi)^{-d/2}\frac{\ps(\x)}{t}\int_0^t\rmd t'(Dt'+h^2)^{-d/2}\nonumber\\
&=(4\pi)^{-d/2}\frac{\ps(\x)}{t}\times\begin{cases} - \frac{h^{2 - d}}{D \left(1 - \frac{d}{2}\right)} + \frac{\left(D + h^{2}\right)^{1 - \frac{d}{2}}}{D \left(1 - \frac{d}{2}\right)} & \text{for}\: d \neq 2 \nonumber\\- \frac{\log{\left(h^{2} \right)}}{D} + \frac{\log{\left(D + h^{2} \right)}}{D} & \text{otherwise} \end{cases}\nonumber\\
&\simeq(4\pi)^{-d/2}\frac{2\ps(\x)}{Dt}\times\begin{cases} \frac{h^{2 - d}}{d - 2} & \text{for}\: d > 2 \\- \log{\left(h \right)} & \text{for}\: d = 2 \end{cases}\ .\label{divergence_return_integral} 
\end{align}

This gives for Gaussian $U$ with width $h$ the result
\begin{align}
\text{var}^\x_\rho(t)\overset{h\to 0}{\simeq}(4\pi)^{-d/2}\frac{4\ps(\x)}{Dt}\times\begin{cases} \frac{h^{2 - d}}{d - 2} & \text{for}\: d > 2 \\- \log{\left(h \right)} & \text{for}\: d = 2 \end{cases} ,
\label{add}
\end{align}
where only the numerical prefactor changes if we choose other indicator functions, since the relevant part (close to $\x$) of any finite size window function can be bounded from above and below by Gaussian window functions.

To extend to general diffusion matrices $\f D(\z)$, we first
  note that for $h\to 0$ only the \emph{local} diffusion matrix $\f
  D(\x)$ at position $\x$ will enter the result, and, if the local $\f D(\x)$ is not isotropic we transform to coordinates where the
diffusion matrix is diagonal, $\f D(\x)={\rm
  diag}(D_1(\x),\dots,D_d(\x))$. One can check this
  by Taylor
  expanding around $\x$ in $h$ in the local coordinate frame, isolating the leading order term, keeping in mind that $\f
  D(\z)$ was assumed to be smooth. In the local coordinates we then
  need to evaluate the integral
\begin{align}
\int_0^t\rmd t' \prod_{i=1}^d(D_i(\x)t'+h^2)^{-1/2},
\end{align}
whose integrand 
can be bounded by
\begin{align}
(\max_j(D_j(\x))t'+h^2)^{-1/2}\le (D_i(\x)t'+h^2)^{-1/2} \le (\min_j(D_j(\x))t'+h^2)^{-1/2},
\end{align}
implying that in the final result $D$ in Eq.~\eqref{add} can be
replaced by $\tilde D(\x)\in[\min(D_i(\x)),\max(D_i(\x))]$, 
\begin{align}
\text{var}^\x_\rho(t)\overset{h\to 0}{\simeq}(4\pi)^{-d/2}\frac{4\ps(\x)}{\tilde D(\x)t}\times\begin{cases} \frac{h^{2 - d}}{d - 2} & \text{for}\: d > 2 \\- \log{\left(h \right)} & \text{for}\: d = 2 \end{cases}.\label{SM bound local time}
\end{align}
The entries $D_i(\x)$ of the diagonalized $\f D(\x)$ are the eigenvalues, hence in general $\tilde D(\x)\in[\lambda(\x)_{\rm min},\lambda(\x)_{\rm max}]$ is bounded by the lowest and highest eigenvalues $\lambda(\x)_{\rm min}$ and $\lambda(\x)_{\rm max}$ of the matrix $\f D(\x)$. This proves the density variance result in Eq.~\eqref{bounds h to 0}.

\subsection{Correlation of current and density}
Now consider the small-window limit for correlations with $\x=\y$
defined as $\f C_{\f J\rho}^{\x\x}(t)\equiv\E{\overline{\f J^U_\x}(t)\overline{\rho^U_\x}(t)}_{\rm s}-\E{\overline{\f J^U_\x}(t)}_{\rm s}\E{\overline{\rho^U_\x}(t)}_{\rm s}$,
given according to Eq.~\eqref{current-density-result} by
\begin{align}
\f C_{\f J\rho}^{\x\x}(t)=&\integrals^{t,U}_{\x\x}\left[\bj_\z P_{\z'}(\z,t')+\revj_\z P_{\z}(\z',t')-2\js(\z)\ps(\z')\right].
\end{align}
Recall that $\bj_\z=\f F(\z)-\f D(\z)\nabla_{\z}$. As
always the term involving the mean values is finite for $h\to 0$. We first look at $\integrals^{t,U}_{\x\x}[\bj_\z P_{\z'}(\z,t')]$, again first for constant isotropic diffusion $\f D(\z)=D\f 1$.

Here we have to use $G_{\rm short,2}$, see Eq.~\eqref{short time propagators}, since the integrals over $G_{\rm short,3}$ vanish. Hence consider
\begin{align}
\integrals^{t,U}_{\x\x}[\bj_\z P_{\z'}(\z,t')]&=\frac{1}{t}\int_0^t\rmd t'\left (1-\frac{t'}{t}\right )\int\rmd^dz\int\rmd^dz'U^h_\x(\z)U^h_\x(\z')\bj_\z P_{\z'}(\z,t')\nonumber\\
&\simeq\frac{\ps(\x)}{t}\int_0^t\rmd t'\int\rmd^dz\int\rmd^dz' U^h_\x(\z)U^h_\x(\z')[\f F(\z)-D\nabla_{\z}]G_{\rm short,2}(\z,t'|\z'),
\end{align}
where we can use $\integrals^{t,U}_{\x\x}[\f F(\z)P_{\z'}(\z,t')]\simeq\f F(\x)\integrals^{t,U}_{\x\x}[P_{\z'}(\z,t')]\simeq\f F(\x)\times\eqref{divergence_return_integral}$ and we compute
\begin{align}
&\nabla_{\z}G_{\rm short,2}(\z,t'|\z')\nonumber\\
&=(4\pi Dt')^{-d/2}\left[1+\frac{1}{2D}(\z-\z')\cdot\f F(\z')\right]\nabla_{\z}\exp\left[-\frac{\left [\z-\z'\right]^2}{4Dt'}\right]+(4\pi Dt')^{-d/2}\frac{\f F(\z')}{2D}\exp\left[-\frac{\left [\z-\z'\right]^2}{4Dt'}\right]\nonumber\\
&=-(4\pi Dt')^{-d/2}\left[1+\frac{1}{2D}(\z-\z')\cdot\f F(\z')\right]\frac{\z-\z'}{2Dt'}\exp\left[-\frac{\left [\z-\z'\right]^2}{4Dt'}\right]+(4\pi Dt')^{-d/2}\frac{\f F(\z')}{2D}\exp\left[-\frac{\left [\z-\z'\right]^2}{4Dt'}\right].
\end{align}
By symmetry, the spatial integrals over $(\z-\z')\exp\left[-\frac{\left [\z-\z'\right]^2}{4Dt'}\right]$ vanish and we are left to compute
\begin{align}
&-D\int\rmd^dz \int\rmd^dz' U^h_\x(\z)U^h_\x(\z')\nabla_{\z}G_{\rm short,2}(\z,t'|\z')\nonumber\\
&\simeq -D(4\pi Dt')^{-d/2}\int\rmd^dx \int\rmd^dy U(\x)U(\y)\left (-\frac{[(\x-\y)\cdot\f F(\y)](\x-\y)}{4D^2t'}+\frac{\f F(\z')}{2D}\right )\exp\left[-\frac{\left [\z-\z'\right]^2}{4Dt'}\right],
\end{align}
where the second term gives $-\frac{1}{2}\f
F(\x)\integrals^{t,U}_{\x\x}[P_{\z'}(\z,t')]$. The remaining term,
noting that $\f F(\z')\simeq\f F(\x)$ and integrating out all directions except $k$ for the $F_k(\x)$ component (by symmetry $(z_i-z'_i)(z_j-z'_j)$ integrates to zero if $i\ne j$), becomes
\begin{align}
&\frac{(4\pi Dt')^{-d/2}}{4Dt'}\int\rmd^dz\int\rmd^dz' U^h_\x(\z)U^h_\x(\z')[(\z-\z')\cdot\f F(\x)](\z-\z')\exp\left[-\frac{\left [\z-\z'\right]^2}{4Dt'}\right]\nonumber\\
&=\frac{\f F(\x)}{4Dt'}\int\rmd z_1 \int\rmd z'_1 U^1(z_1)U^1(z'_1)(z_1-z'_1)^2G_{\rm short,3,one-dim}(z_1,t' |z'_1)\nonumber\\
&=\frac{\f F(\x)}{4Dt'}\frac{D h^{2} t'}{\sqrt{\pi} \left(D t' + h^{2}\right)^{\frac{3}{2}}}\nonumber\\
&=\frac{\f F(\x)h^{2}}{4\sqrt{\pi} \left(D t' + h^{2}\right)^{\frac{3}{2}}}.
\end{align}
This term is subdominant as we see from the time integral
\begin{align}
\integrals^{t,U}_{\x\x}[-D\nabla_{\z} P_{\z'}(\z,t')]&\simeq-\frac{\f F(\x)}{2}\integrals^{t,U}_{\x\x}[P_{\z'}(\z,t')]+
\frac{\ps(\x)\f F(\x)h^2}{4\sqrt{\pi}t}\underbrace{\int_0^t\rmd t'\left(D t' + h^{2}\right)^{-\frac{3}{2}}}_{h^{-1}}\nonumber\\
&\simeq-\frac{\f F(\x)}{2}\integrals^{t,U}_{\x\x}[P_{\z'}(\z,t')].
\end{align}
Hence, overall we get
\begin{align}
\integrals^{t,U}_{\x\x}[\bj_\z P_{\z'}(\z,t')]&=\frac{\f F(\x)}{2}\integrals^{t,U}_{\x\x}[P_{\z'}(\z,t')].\label{corr_int_1} 
\end{align}
The generalization to non-constant or non-isotropic $\f D(\z)$ only changes $\integrals^{t,U}_{\x\x}[P_{\z'}(\z,t')]$ but Eq.~\eqref{corr_int_1} is retained.

Now consider $\integrals^{t,U}_{\x\x}[\revj_\z P_{\z}(\z',t')]$. Since this involves derivatives of both $G$ and $\ps$ (at the initial point) we instead take the form $\revj_\z=\js(\z)/\ps(\z)+D\ps(\z)\nabla_{\z}\ps(\z)^{-1}$ such that $\revj_\z P_{\z}(\z',t')=[\js(\z)+D\ps(\z)\nabla_{\z}]G(\z',t'|\z)$ giving
\begin{align}
\integrals^{t,U}_{\x\x}[\revj_\z P_{\z}(\z',t')]&=\frac{1}{t}\int_0^t\rmd t'\left (1-\frac{t'}{t}\right )\int\rmd^dz\int\rmd^dz' U^h_\x(\z)U^h_\x(\z')[\js(\z)+D\ps(\z)\nabla_{\z}]G(\z',t'|\z)\nonumber\\
&\simeq\frac{\js(\x)}{\ps(\x)}\integrals^{t,U}_{\x\x}[P_{\z}(\z',t')]+
D\ps(\x)\int_0^t\rmd t'\int\rmd^dz\int\rmd^dz' U^h_\x(\z)U^h_\x(\z')\nabla_{\z}G_{\rm short,2}(\z',t'|\z),\label{cor_bound1}
\end{align}
where (note that $G_{\rm short,3}(\z,t'|\z')=G_{\rm short,3}(\z',t'|\z)$)
\begin{align}
&\nabla_{\z}G_{\rm short,2}(\z',t'|\z)\nonumber\\
&=(4\pi Dt')^{-d/2}\left[1+\frac{1}{2D}(\z'-\z)\cdot\f F(\z')\right]\nabla_{\z}\exp\left[-\frac{\left [\z-\z'\right]^2}{4Dt'}\right]+(4\pi Dt')^{-d/2}\frac{-\f F(\z)}{2D}\exp\left[-\frac{\left [\z-\z'\right]^2}{4Dt'}\right]\nonumber\\
&=-(4\pi Dt')^{-d/2}\left[1+\frac{1}{2D}(\z-\z')\cdot\f F(\z')\right]\frac{\z-\z'}{2Dt'}\exp\left[-\frac{\left [\z-\z'\right]^2}{4Dt'}\right]-\frac{\f F(\z')}{2D}G_{\rm short,3}(\z',t'|\z).
\end{align}
As before, asymptotically only the last term contributes, giving 
\begin{align}
\integrals^{t,U}_{\x\x}[\revj_\z P_{\z}(\z',t')]&\simeq\frac{\js(\x)}{\ps(\x)}\integrals^{t,U}_{\x\x}[P_{\z}(\z',t')]+
D\ps(\x)\int_0^t\rmd t'\int\rmd^dz\int\rmd^dz' U^h_\x(\z)U^h_\x(\z')\nabla_{\z}G_{\rm short,2}(\z',t'|\z)\nonumber\\
&\simeq\frac{\js(\x)}{\ps(\x)}\integrals^{t,U}_{\x\x}[P_{\z}(\z',t')]+
D\ps(\x)\int_0^t\rmd t'\int\rmd^dz\int\rmd^dz' U^h_\x(\z)U^h_\x(\z')\frac{-\f F(\z')}{2D}G_{\rm short,3}(\z',t'|\z)\nonumber\\
&\simeq\left [\frac{\js(\x)}{\ps(\x)}-\frac{\f F(\x)}{2}\right]\integrals^{t,U}_{\x\x}[P_{\z}(\z',t')].\label{cor_bound2}
\end{align}
Overall, this gives for the correlations (having the same form for anisotropic diffusion)
\begin{align}
\f C_{\f J\rho}^{\x\x}(t)\overset{h\to 0}\simeq\frac{\js(\x)}{\ps(\x)}\integrals^{t,U}_{\x\x}[P_{\z}(\z',t')]\simeq\frac{\js(\x)}{2\ps(\x)}\text{var}^\x_\rho(t).\label{cor_bound_full} 
\end{align}
This proves the correlation result in Eq.~\eqref{bounds h to 0}.

\subsection{Current variance}
We now turn to the current variance, see Eq.~\eqref{current covariance result} for $\x=\y$,
\begin{align}
\text{var}^{\x}_{\f J}(t)=\frac{2}{t}\int\rmd^d z{\rm Tr}\f D(\z)U^h_\x(\z)U^h_\x(\z)\ps(\z)+2\integrals^{t,U}_{\x\x}\left[\bj_\z\cdot\revj_{\z'} P_{\z'}(\z,t')-\js(\z)\cdot\js(\z')\right].
\end{align}
The first term for $h\to 0$ gives
\begin{align}
\frac{2{\rm Tr}\f D(\z)}{t}\int\rmd^d xU^h_\x(\z)U^h_\x(\z)\ps(\z)\simeq\frac{2{\rm Tr}\f D(\x)}{t}\ps(\x)U^h_\x(\x),
\end{align}
where $U^h_\x(\x)\propto h^{-d}$ is the height of the delta-function approximation, e.g.\ $U^h_\x(\x)=(2\pi)^{-d/2}h^{-d}$ for Gaussian $U^h_\x$. In the derivation (see Sec.~\ref{derivation}) this term occurred from cross correlations $\rmd W_{t_1}\rmd W_{t_1}=\rmd t'\ne 0$ in the noise part, hence it can be seen to come from zero time-differences $t'=t_2-t_1=0$. Such a term does not appear in the density variance or density-current correlation since there $\rmd t_1\rmd t_2=0$ and $\rmd t_1\rmd W_{t_2}=0$ would occur instead of $\rmd W_{t_1}\rmd W_{t_2}$.

Due to the fast $h^{-d}$ divergence, the dominant limit does not
depend on terms with no or only one derivative since they were shown
to scale at most as $h^{2-d}$. The only new term is the second
derivative for which we see that
\begin{align}
&\int_0^t\rmd t'D\nabla_{\z}\cdot(-\nabla_{\z'})G_\text{short,3}(\z,t'|\z')\nonumber\\
&=\int_0^t D\nabla_{\z}^2G_\text{short,3}(\z,t'|\z')\nonumber\\
&=\int_0^t\rmd t'\partial_{t'}G_\text{short,3}(\z,t'|\z')\nonumber\\
&=\left [G_\text{short,3}(\z,t'|\z')\right ]_0^t\nonumber\\
&=G_\text{short,3}(\z,t|\z')-\delta(\z-\z'),
\end{align}
such that
\begin{align}
\integrals^{t,U}_{\x\x}\left[\bj_\z\cdot\revj_{\z'} P_{\z'}(\z,t')\right ]
&\simeq -D^2\frac{\ps(\x)}{t}\int_0^t\rmd t'\int\rmd^dz\int\rmd^dz'U^h_\x(\z)U^h(\z)\nabla_{\z}\cdot\nabla_{\z'}G_\text{short,3}(\z,t'|\z')\nonumber\\
&\simeq -D\frac{\ps(\x)}{t}\int\rmd^dz\int\rmd^dz'U^h_\x(\z)U^h(\z')\delta(\z-\z')\nonumber\\
&\simeq -D\frac{\ps(\x)}{t}U^h(\x).
\end{align}
For non-isotropic and possibly non-constant $\f D(\z)\ne D\f 1$,
  we again note that for $h\to 0$ only $\f D(\z)$ at $\z=\x$ matters,
  and move to the basis where $\f D=\f D(\x)$ is diagonal, where we have
\begin{align}
D^2\nabla_{\z}\cdot\nabla_{\z'}\longrightarrow\sum_{i=1}^d D_i^2 \partial_{z_i}\partial_{z_{i'}}.
\end{align}
The operator we need is $\nabla_{\z}\f D\nabla_{\z'}=\sum_i D_i\partial_{z_i}\partial_{z_{i'}}$ so we bound one of the $D_i$ in $D_i^2$ by $D'\in[\min(D_i),\max(D_i)]$ such that we get
\begin{align}
\integrals^{t,U}_{\x\x}\left[\bj_\z\cdot\revj_{\z'} P_{\z'}(\z,t')\right ]
&\simeq -D'\frac{\ps(\x)}{t}U^h(\x).
\end{align}
Since ${\rm Tr}\f D=\sum_i D_i$ we have $\tilde{D}'\equiv\frac{{\rm Tr}\f D-D'}{d-1}\in[\min(D_i),\max(D_i)]$ and we can write
\begin{align}
\text{var}^{\x}_{\f J}(t)&\simeq \frac{2{\rm Tr}\f D}{t}\ps(\x)U^h(\x)-2D'\frac{\ps(\x)}{t}U^h(\x)=\frac{2\tilde{D}'}{t}\ps(\x)(d-1)U(\x),\label{bound_var_j} 
\end{align}
where $U(\x)\propto h^{-d}$. This proves the current variance result in Eq.~\eqref{bounds h to 0}. Thus, we see that the current fluctuations diverge for $h\to 0$, except in one-dimensional space where $d-1=0$.

\subsection{Limit of no coarse graining in the one-dimensional case}\label{1d h=0}
In the one-dimensional case, the variance of empirical density and
current remain finite for $h\to 0$ which allows to take the limit to
$U^{h=0}_x(x')=\delta(x-x')$. In terms of the stochastic integrals, the
one-dimensional case is much simpler, since any one-dimensional
function $U^h_x(x')$ possesses an antiderivative -- a primitive function
$\mathcal U^h_x(x')=\int^{x'}U^h_x(x'')\rmd x''$ such that
$U^h_x(x')=\partial_{x'}\mathcal{U}_x(x').$ This implies for the
Stratonovich integral that 
\begin{align}
t\overline{J^U_x}(t)=\int_0^t U^h_x(x_\tau)\circ\rmd x_\tau=\mathcal{U}_x(x_t)-\mathcal{U}_x(x_0).
\label{1d current primitive}
\end{align}
Thus, the stochastic current is no longer a functional but only a function of the initial- and end-point of the trajectory. Its moments are directly accessible, e.g.\
\begin{align}
\E{[\overline{J^U_x}(t)]^2}_{\rm s}=\frac{1}{t^2}\E{\left [\mathcal{U}_x(x_t)-\mathcal{U}_x(x_0)\right ]^2}_{\rm s}=\frac{1}{t^2}\int\rmd z\int\rmd z'\left [\mathcal{U}_x(z)-\mathcal{U}_x(z')\right ]^2P_{z'}(z,t).
\end{align}
If $U$ is Gaussian, then $\mathcal{U}_x$ is the error function such
that $\left[\mathcal{U}_x(x)-\mathcal{U}_x(y)\right]^2\le 1$ and thus
$\E{[\overline{J^U_x}(t)]^2}_{\rm s}\le 1/t^2$. This also holds in the
limit of a delta function where the primitive function becomes a
Heaviside step function and we get that the current can only be $0$ or
$\pm t^{-2}$, see Fig.~\ref{Fg1d}. The current defined with a
delta function at $x$ simply counts the net number of crossing through
$x$ such that all crossings except maybe one cancel out. 
Note that the reasoning above only holds for the current defined with a Stratonovich integral---the same definition with an It\^o or anti-It\^o integral would give a divergent current for the delta function.

To obtain a
$1/t$-term as in large deviations one would need to have a
steady-state current which could e.g.\ be achieved by generalizing to
periodic boundary conditions. Then the current would depend on the initial and final point and, in addition, also on the net number of crossings of the full interval between the boundaries of the system.

Fig.~\ref{Fg1d} shows the time-integrated density and current, i.e.\ the empirical density and current Eq.~\eqref{def_current} multiplied by the total time $t$. Fluctuations remain in the same order of magnitude for $h\to 0$ (see Fig.~\ref{Fg1d}c,e). We see that the time-integrated current is bounded by $1$ which is due to the fact that it simply counts the net number of crossings. According to Eq.~\eqref{1d current primitive} it only depends on the initial-point $x_0$ and end-point $x_t$, in this case $x_{10}$.
\begin{figure}[!ht]
\begin{center}
\includegraphics[scale=1]{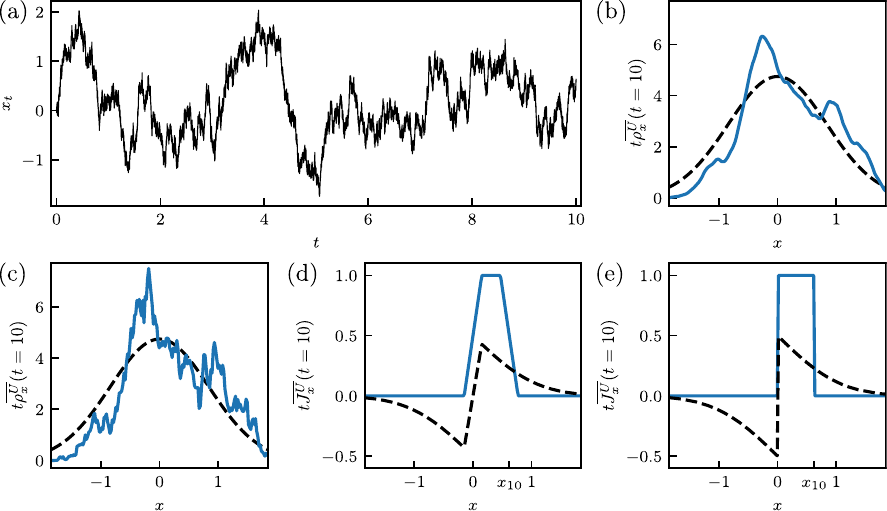}
\caption{\label{Fg1d}\blue{{(a)} One-dimensional Brownian motion in a harmonic potential (see Eq.~\eqref{OUP} in 1d with parameters $r=\sqrt{2}$ and $D=1$) starting at $x_0=0$ and ending at $x_{10}=0.62$. {(b)} Time-integrated density 
of the trajectory in (a) as a function of $x$ for normalized window function $U^h_x(x')=h^{-1}\mathbbm 1_{\abs{x-x'}\le h/2}$ with width $h=0.3$. The dashed line shows the expectation value of the time-integrated density conditioned on $\x_0=0$. {(c)} As in (b) with width $h=0.001$. {(d)} Time-integrated current for window as in (b) width $h=0.3$. The dashed line shows the expectation value of the time-integrated current conditioned on $\x_0=0$. {(e)} As in (d) for width $h=0.001$.}}
\end{center}\end{figure}}

\bibliography{bibPRLnew.bib}
\end{document}